\documentclass[onecolumn,pre,amsmath,amssymb,showpacs,superscriptaddress]{revtex4}
\usepackage{graphics}
\usepackage{graphicx}
\usepackage{subfigure}
\usepackage{psfrag}
\usepackage{color}
\usepackage{bm}
\usepackage{natbib}

\def\be{\begin{equation}}
\def\ee{\end{equation}}

\def\be{\begin{array}{l}}
\def\ee{\end{array}}
\def\eem{\end{array}}
\def\ben{\begin{eqnarray}}
\def\een{\end{eqnarray}}
\def\bee{\begin{eqnarray*}}
\def\eee{\end{eqnarray*}}

\begin{document}

\title{A Paradigm for Modeling and Computation of Gas Dynamics}

\author{Kun Xu}
\email[Email:]{makxu@ust.hk}
\affiliation{Department of Mathematics, Department of Mechanical and Aerospace Engineering, Hong Kong University of Science and Technology, Hong Kong}
\author{Chang Liu}
\email[Email:]{cliuaa@connect.ust.hk}
\affiliation{Department of Mathematics, Hong Kong University of Science and Technology, Hong Kong}

\begin{abstract}

In the continuum flow regime, the Navier-Stokes equations are usually used for the description of gas dynamics.
On the other hand, the Boltzmann equation is applied for the rarefied gas dynamics.
Both equations are constructed from modeling flow physics in different scales.
Fortunately, due to the distinct separation of scales, i.e., the hydrodynamic and kinetic ones, both Navier-Stokes equations and the Boltzmann equation are valid in their
respectable domains.
However, in real physical application, there may not have such a distinctive scale separation.
For example, around a hypersonic flying vehicle, the flow physics at different regions may correspond
to different regimes, where the local Knudsen number can be changed in several order of magnitudes.
With a variation of modeling scale, theoretically a continuous governing equation from kinetic Boltzmann equation to the hydrodynamic Navier-Stokes
equations should exist.
However, due to the difficulties of a direct modeling of flow physics in the scale between the kinetic and hydrodynamic ones,
there is basically no reliable theory or  valid governing equation to cover the whole transition regime.
In fact, it is  an unresolved problem about the exact scale for the validity of the NS equations as Reynolds number decreases.
The traditional computational fluid dynamics (CFD) is based on the numerical solution of partial differential equations (PDE), and it targets on the recovering of the
 exact solution  of the PDEs as mesh size and time step converging to zero.
 This methodology can be hardly applied here because there is no such a complete
 PDE for flow physics in all scales.
It may be the time to combine the modeling and computation together without going through the process of constructing PDEs.
In other words, the CFD research is not only to obtain the numerical solution of governing equation,
but also to construct a valid discrete governing equation to identify the flow physics in the mesh size and time step scales.
In this paper, we are going to present the idea for the modeling and computation.
This methodology leads to the unified gas-kinetic scheme (UGKS) for flow simulation in all flow regimes. Based on UGKS, the
boundary for the validation of the
Navier-Stokes equations can be quantitatively evaluated.
The combination of modeling and computation provides a paradigm  for the description of multiscale transport process.

\end{abstract}


\maketitle


\section{Introduction}

\begin{figure}[!htb]
\centering
\includegraphics[bb=291 118 667 438, clip,width=8cm]{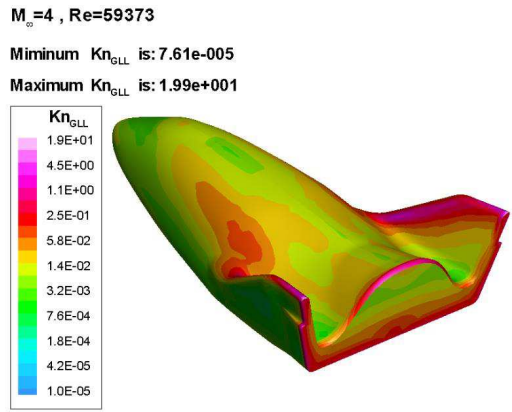}
\caption{Local Knudsen number around a flying vehicle at Mach number $4$ and Reynolds number $59,373$ calculated by D.W. Jiang using a unified gas kinetic scheme \cite{jiang}.}
\label{fig:vehicle}
\end{figure}

The theoretical study of gas dynamics is mostly based on the analytical and numerical solutions of the Euler and Navier-Stokes (NS) equations.
 The modeling of the NS is coming from the conservative physical laws, and the inclusion of additional constitutive relationships
 for the stress and strain, and the relationship between the heat flux and temperature gradient.
But, the scale for the validity of the NS equations has never been clearly defined, even though it always refers as the hydrodynamic one.
Even with the wide applications of the NS equations, the boundary for the validity of NS equations is unclear.
For a hypersonic flow around a flying vehicle, different flow physics may emerge at different regions, such as the highly non-equilibrium shock layer,
low density trailing edge, and the wake turbulence.
Fig. \ref{fig:vehicle} presents the local Knudsen number around a flying vehicle at Mach number $4$ and Reynolds number $59,373$.
As shown in this figure, the local Knudsen number can cover a wide range of values with five order of magnitude difference.
It seems that a single scale governing equation can be hardly applicable in an efficient way to all flow regimes.
On the other hand, the Boltzmann equation is derived on a well-defined modeling scale, which is the particle mean free path and the particle
mean traveling time between collisions \cite{chapman}. This is also the finest resolution of the Boltzmann equation.
Only under such a modeling scale, the particle transport and collision can be separately formulated.
In the kinetic scale, the particle distribution  is modeled as a field, and the Boltzmann equation
becomes a statistical modeling equation. The Boltzmann equation can be numerically solved through the Direct Simulation Monte Carlo (DSMC) method \cite{bird}
or the direct Boltzmann solver \cite{aristov} with the numerical resolution on the same scale.
Since the Boltzmann equation has a much refined resolution in comparison with hydrodynamic one,
in order to derive the NS equations a coarse-graining process has to be
used. One of the most successful theoretical study of the Boltzmann equation is the Chapman-Enskog expansion,
where with a proper stretching of the space and time scales the NS equations can be obtained.
It is fortunate that due to the separation of scales between NS and Boltzmann equations, both equations can be confidently used in their respective scales.
Even though the NS equations can be correctly derived form the Boltzmann equation, there are tremendous difficulties
to derive other equations between the kinetic and hydrodynamic scales,
which span over the whole non-equilibrium flow regime.
The difficulties are associated with the following reasons.
Firstly,  how to define  a continuously varying scale between the kinetic and hydrodynamic ones to do the modeling and derive the equations.
Secondly, what kinds of flow variables can be used to describe the flow motion between these two limits.
Thirdly, in the transition region there is no clear scale separation and the conventional mathematical tool may not be applicable.
For the NS equations, there are only five flow variables,
such as mass, momentum, and energy, to describe the dynamics \cite{landau}.
However, for the Boltzmann equation, there are theoretically infinite number degrees of freedom due to the
capturing of individual particle movement.
How many flow variables should be properly used between these two limiting cases  to recover all possible non-equilibrium state are basically unclear.
All extended thermodynamic theories or irreversible thermodynamics are focusing on the study of flow close to equilibrium only.
In fact, we have no much knowledge about the non-equilibrium physics between the hydrodynamic and kinetic scales.

In reality, for the gas dynamics the use of  distinct governing equations, such as NS and Boltzmann, is based
on their distinct scales and these descriptions are incomplete.
With the variation of the modeling scale, there should exist a continuous spectrum of dynamics between these two limits.
The multiple scale  equation is needed to capture the scale-dependent flow physics from the kinetic to the hydrodynamic ones.
With great difficulty by choosing an appropriate modeling scale in the theoretical study,
the  computation provides us an opportunity to do direct modeling with a freely varying scale, which is the mesh size and time step.
In other words, the traditional derivation of governing equation can be replaced by a modeling procedure in a discretized space.
Therefore, the numerical algorithm itself provides  governing equation for the description of gas dynamics.
Based on the direct modeling on the mesh size and time step, a unified gas-kinetic scheme (UGKS) has been developed for the flow description in all regimes \cite{xu2010,xu-book}.
The main purpose of this paper is to point out the way beyond the traditional numerical PDE methodology
to the construct the direct modeling method.
At the same time,
 we are going to use the direct modeling to validate the NS equations through case studies.
The paradigm for modeling and computation is useful in the study of multiple scale transport process.

\section{Gas Dynamics Modeling}

\begin{figure}[!htb]
\centering
\includegraphics[bb=90 207 694 387, clip,width=13cm]{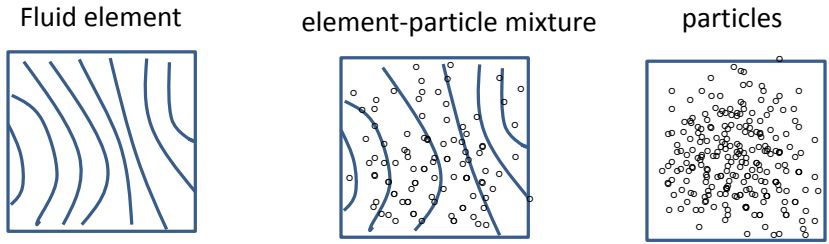}
\caption{Modeling variations from the fluid element in hydrodynamic scale (left) to the particle representation in kinetic scale (right) through the non-equilibrium
regime with a variable degree of freedom (middle).}
\label{fig:wave-particle}
\end{figure}

Now consider a box with the length scale $L$, such as with a value $L=0.01m$, and the box is supposed to hold different amount of molecules, see Fig.{\ref{fig:wave-particle}}.
Under the standard atmospherical condition, the number density of molecules is $n= 2.687 \times 10^{25} m^{-3}$.
In the following mental experiments, we assume that the number density of the particles inside the box can be changed significantly to different levels.
Define the diameter of the molecule as $d$, which is on the order of $d=3.7\times 10^{-10}m$,
and the mean free path between the collisions of the molecule  $l$.
The relationship between $d$ and $l$ is $l = 1/{{\sqrt{2}}(\pi n d^2)}$, such as $6.1 \times 10^{-8} m$ in the standard atmospheric condition.
The density of the gas inside the box can be defined as
$\rho = nm$ and $m$ is molecular mass.
The Kundsen  number defined as ${\mbox{Kn}} = l/L$ indicates the rarefaction of the molecular distribution. The scales from the molecular diameter to the
dimension of the box can be varied across a scale of ten orders of magnitude.

Let's assume a constant gas temperature $T$ and different number density inside the box.
In the kinetic limit, such as at the Knudsen number $Kn \simeq 1$, the molecules can freely move through the box from one end to the other side,
and the interactions between the molecules and the walls are equally important.
As the Knudsen number reduces with the increment of molecules inside the box,
such as $Kn=0.1$, each particle may take $10$ collisions to move from one end to the other end of the box.
At the same time, each particle can still move freely to anywhere inside the box.
There is fully penetration among all molecules.
With the Boltzmann modeling, the flow physics under such a condition can be easily solved using a mesh size on the order of
the particle mean free path.
In the Boltzmann solver, particle free transport and collision can be treated separately. If there is no bulk velocity for the molecules inside the box,
for such a system with $  0.1 < Kn < 10   $  information inside the box can propagate from one end to another end through molecular motion
with a speed $C_r \sim \sqrt{RT}$, where $R=k/m$ is the gas constant and $k$ is the Boltzmann constant,  as shown in the right sub-figure in Fig.{\ref{fig:wave-particle}}.

Before we consider the syetem with a continuous reduction of Knudsen number, let's go to another limit, i.e., the hydrodynamic one.
Under such a situation, such as at the standard atmospheric condition there are an order of $10^{19}$ particles inside the box.
At such a limit, the Knudsen number can reach an extremely small  value, as low as $10^{-6}$.
If we still use the Boltzmann modeling to study the system, we need a high resolution calculation with the mesh size on the
order of $10^{-8}m$.
For such a system, due to the high molecular density and small particle mean free path, in order to study the system efficiently, the fluid element modeling can be used
and  the traditional hydrodynamic equations are accurate enough for the description of flow structure in such a large scale, such as at the resolution
of a cell size $10^{-4} m$.
With the intensive collisions in the scales $10^{-8} m$ and $10^{-10} s$ for the particle collision,
the exchange of the momentum and energy will equalize the temperature and averaged velocity locally. Therefore, the local
equilibrium assumption can be achieved in the hydrodynamic scale $10^{-4} m$.
With the separation of the scales in the hydrodynamic and kinetic ones,
the NS scale modeling for such a system can use the fluid element concept, where the
molecules inside the box can be separated into different distinguishable elements with a gigantic amount of molecules inside each unit.
Between the elements, there is pressure, viscous friction, and heat exchange,
but there is no mass or molecules penetration between the elements due to scale separation, such as the absence of mass diffusion term.
The interactions between fluid elements are basically through waves, such as the left sub-figure in Fig.{\ref{fig:wave-particle}}.
This is also the foundation for using the equation of state to each isolated fluid element through the classical thermodynamics.
In other words, in the continuum NS limit, the intensive particle collision prevents the particle from free penetration
between elements. The energy exchange, such as the work done by the force and the heat transfer, takes place
through the boundary, such as the heat diffusion in the Fourier's law.
In such a case, any information in the gas system will propagate through wave behavior, i.e., a process for each fluid element to push neighboring one.
This wave propagating process has the same speed as the molecular motion $C_c \sim \sqrt{RT}$ in the continuum limit.
Only under the fluid element picture, there have Lagrangian and Euler descriptions for the gas dynamics.
The fluid element picture sometimes is associated with  difficulties to cope with other requirement, such as the non-slip boundary condition.
Under such a condition, a fluid element needs to be stretched forever in the boundary layer, which cannot be true.
More importantly, for the NS equations there is no a clear definition of the scale for the validity of the equation itself, such as the
scale of element where the constitutive relationship can be faithfully applied for the gas dynamic equations.

Starting from the continuum limit, if the gas density inside the box is reduced,
the mean free path of the gas molecule will increase.
The assumption of the isolated fluid element will break down as the particle penetration effect emerges. With the further reduction of the gas density,
the fluid element assumption has to be abandoned due to the intensive particle exchange among neighboring fluid elements.
During this process, both the pressure interaction (waves through fluid element) and particle penetration (particles free transport)
will take effect, such as the middle figure in Fig.{\ref{fig:wave-particle}}.
In this regime, the information can propagate through the wave interaction and particle transport, which have the same speed of  $C_m \sim \sqrt{RT}$.
In terms of physical modeling, in such a scale it is difficult to give a complete description of the flow system using fluid element picture or individual particle motion.
Unfortunately, all extended hydrodynamic equations or moment equations derived from the Boltzmann equation
are intrinsically based on the fluid
element assumption through macroscopic flow variables.
To reach a macroscopic level of description, starting from the microscopic reality a coarse graining process has to be used. During this process,
a certain amount of information  gets lost and a corresponding uncertainty is added to the macroscopic description, such as the supplement of constitutive relationship.

With the reduction of gas density,
the degrees of freedom for an accurate description of the flow system increases continuously from the hydrodynamic to the
kinetic level.
In other words, the construction of the extended hydrodynamic equations with a fixed number of flow variables, such as Burnett, Super-Burnett, or moments equations,
cannot give  a complete representation.
On the other hand, with a kinetic scale resolution everywhere, the direct use of the Boltzmann equation will be very expensive
to present the solution with an enforced mean free path scale resolution.
In fact, in the regime between the hydrodynamic and kinetic ones,
there is basically no any valid governing equation yet with a variation of degrees of freedom for the capturing of non-equilibrium effect.
Furthermore, no proper flow variables can be identified to give a valid description for such a system, such as the mathematical description
in scale with multiple particle collisions.
Therefore, this regime is basically unexplored even though we have two  successful limiting governing equations, i.e., NS and Boltzmann, in two distinct and separate modeling scales.
The inseparable or continuous variation of scales in the transition regime makes theoretical modeling difficult.
For example, how could we present a mathematical formulation in a scale with a few particle mean free path?
 The separation of mathematical formulation of particle free transport and collision in the Boltzmann modeling cannot be
applied in such a coarse graining scale with multiple collisions for individual particle. Even though it is difficult to formulate a mathematical description in such a scale,
it can be done directly through the modeling of the gas evolution process in the mesh size scale. This is the basic idea for the modeling and computation together.

In order to give a full description of gas dynamics at different scales,
we have to construct valid governing equations for a continuous variation of flow physics.
The scale used for the modeling can be defined as the ratio of the mean free path over the cell size, the so-called cell Knudsen number $Kn_c = l/\Delta x$.
The pure theoretical study has difficulty in identifying a modeling scale.
Fortunately, for the computation we can adopt the mesh size as a modeling scale.
Based on the direct modeling on the mesh size scale,
a continuous description of flow physics has been obtained
in the unified gas-kinetic scheme (UGKS) \cite{xu2010,xu-book}, which covers the NS and Boltzmann physics in the limiting cases,
and provides a valid solution in the transition regime \cite{liu}.
Without using the direct modeling methodology, it will be difficulty for any numerical PDE approach to properly connect NS and Boltzmann solutions \cite{chen}.

As shown in the later sections,  at low Reynolds number case the numerical computation for the direct NS solver requires a very small time step.
This indicates that the NS modeling is not appropriate here,
where the particle penetration has been ignored in the fluid element assumption.
For UGKS,  a physical time step, which is independent  of Reynolds numbers, can be used uniformly in all flow regimes.
This is consistent with the above analysis, where the physical propagating speed is independent of the gas density.
The aim  of this paper is to figure out the dynamics differences quantitatively between the
UGKS and NS modeling, and points out the importance to adopt direct modeling for computation in order to solve the multiple scale transport problem.

\section{Distinct governing equations and direct modeling scheme}

In order to present the ideas of the conventional CFD simulation and the direct modeling approach, we are going to
use  the linear advection diffusion equation and the kinetic Boltzmann BGK model for flow description
in different scales. The extension of the scheme to gas dynamic equations will be presented in the next section as well.

\subsection{Hydrodynamic equation and its connection to kinetic model equation}

The kinetic Boltzmann equation and the hydrodynamic Navier-Stokes equations are obtained based on different modeling scales.
In the kinetic mean free path scale, the BGK equation models the flow physics as \cite{bgk}
\begin{equation}\label{bgk}
  \partial_t f+c\partial_xf =\frac{1}{\tau}(g-f),
\end{equation}
for the evolution of a gas distribution function $f$ with free transport (left) and collision term (right) effects.
In Eq.\eqref{bgk}, $f(c,x,t)$ is the velocity distribution function, $c$ is the microscopic particle velocity, $g$ is the equilibrium state, and
$\tau$ is the collision time.
In the hydrodynamic scale with the fluid element approximation, the linear advection-diffusion equation is
\begin{equation}\label{advection-diffusion}
  u_t+au_x=\nu u_{xx},
\end{equation}
for the propagation  of macroscopic variable $u$ with macroscopic velocity $a$, and diffusive mechanism with a constant viscosity coefficient $\nu$.

The macroscopic and microscopic quantities are related through
\begin{equation}
  u(x,t)= \int_R f(c,x,t) dc.
\end{equation}
The equilibrium Maxwellian distribution $g$ is
\begin{equation}\label{maxwell}
  g=u\frac{1}{\sqrt{\theta \pi}}\mathrm{e}^{-\frac{(c-a)^2}{\theta}}.
\end{equation}
The $\theta$ corresponds to the temperature, which is related to the spread of particle random velocity.

Integrating the particle velocity on both sides of Eq.\eqref{bgk} gives the macroscopic equation
$$\partial_t u+\partial_x F=0,$$
with the flux
$$F(x,t)=\int_R cf(c,x,t) dc.$$
In the continuum regime, with the underlying physical assumption that the variation of $f$ in the hydrodynamic scale is smooth enough
due to substantial particle collisions, the Chapman-Enskog method gives a solution
\begin{equation}
  f^{(N)}(c,x,t)=\sum^N_{n=0}\tau^n f_n(c,x,t).
\end{equation}
The first order expansion becomes
\begin{equation}\label{CE}
  f^{(1)}(c,x,t)=(u-\tau(c-a)\partial_xu)\frac{1}{\sqrt{\theta \pi}}
  \mathrm{e}^{-\frac{(c-a)^2}{\theta}}.
\end{equation}
The corresponding macroscopic flux reads
\begin{equation}\label{flux-1st}
  F^{(1)}=au-\frac{\theta \tau}{2}\partial_x u,
\end{equation}
which leads to the advection-diffusion
$$u_t+au_x=\frac{\theta \tau}{2} u_{xx}.$$
By comparing the coefficient with equation \eqref{advection-diffusion},
we have the relation
$$\nu=\frac{\theta \tau}{2}.$$
The expansion \eqref{CE} converges when $\tau$ is small, and Eq.\eqref{advection-diffusion} may not be consistent with Eq.\eqref{bgk} when $\tau$ gets large.

\subsection{The direct modeling scheme and the hydrodynamic equation solver}

We consider a discretization of the space-time $\Omega\times[0,T]$
with constant spatial cell size $\Delta x$ and time step $\Delta t$.
The basic numerical method is an explicit finite volume method.
The evolution equation for the macroscopic conservative variable is
\begin{equation}\label{Finite-volume-u}
  U^{n+1}_i=U^{n}_i+\frac{\Delta t}{\Delta x}(F^{n}_{i-\frac12}-F^{n}_{i+\frac12}),
\end{equation}
where $F^{n}_{i+\frac12}$ is the time averaged numerical flux at cell interface, which can be obtained
from the gas distribution there,
\begin{equation}\label{flux-u}
  F^{n}_{i+\frac12}=\frac{1}{\Delta t}\int_0^{\Delta t}\int_{-\infty}^{\infty}
  cf(c,x_{i+\frac12},t)dc dt.
\end{equation}
The above Eq.(\ref{Finite-volume-u}) is the physical conservation law in a discretized space.
The physics to be captured depends on the modeling of the cell interface distribution function in Eq.(\ref{flux-u}).
As analyzed before, in the hydrodynamic scale, the fluid element will be used to model the flux, such as the solver based on Eq.(\ref{advection-diffusion}).
In the kinetic scale, the
particle transport and collision will take effect, and the solution from Eq.(\ref{bgk}) can be used.
In the direct modeling method, the physics to be simulated will depend on the scale of $\Delta x$ and $\Delta t$
in Eq.(\ref{Finite-volume-u})
with respect to the particle mean free path and collision time.

For the hydrodynamic solver,  we only need to update the above macroscopic conservative flow variable.
However, as the gas density reduces, in the transition flow regime the macroscopic flow variable update alone is not enough for the capturing of the
peculiarity of the non-equilibrium property, where more degrees of freedom are needed to follow the flow evolution.
The unified gas-kinetic scheme (UGKS) is based on the evolution of both
macroscopic variable Eq.(\ref{Finite-volume-u})  and the gas distribution function.
The evolution equation for the microscopic velocity distribution function is
\begin{equation}\label{Finite-volume-f}
  f^{n+1}_i=\left(1+\frac{\Delta t}{2\tau}\right)^{-1}
  \left[f^n_i+\frac{\Delta t}{\Delta x}(\tilde{f}_{i-\frac12}-\tilde{f}_{i+\frac12})
  +\frac{\Delta t}{2}\left(\frac{g^{n+1}}{\tau}+\frac{g^{n}-f^n}{\tau}\right)\right],
  \end{equation}
 where $\tilde{f}_{i+\frac12}$ is the time averaged numerical flux for distribution function,
which is calculated by
\begin{equation}\label{flux-f}
\tilde{f}_{i+\frac12}=\frac{1}{\Delta t}\int_0^{\Delta t}c f(c,x_{i+\frac12},t)dt.
\end{equation}
Here we solve the kinetic BGK Eq.(\ref{bgk}). For solving the full Boltzmann equation, similar technique can be applied \cite{liu}.

For the macroscopic equation solver,  Eq.(\ref{Finite-volume-u}) alone  is used for the update of macroscopic flow variable.
For direct modeling method, the UGKS uses both Eq.(\ref{Finite-volume-u}) and Eq.(\ref{Finite-volume-f}) for the update of flow variable and the
gas distribution function.
The flux function for both schemes is based on the same integral solution of Eq.\eqref{bgk} at a cell interface,
\begin{equation}\label{solution}
  f(c,x,t)=\frac{1}{\tau}\int_0^t g(c,x-cs,t-s)\mathrm{e}^{-\frac{s}{t}}ds+
  \mathrm{e}^{-\frac{t}{\tau}}f_0(c,x-ct,0),
\end{equation}
where $f_0$ is the initial condition.
This is a multiple scale transport solution, which covers from the free molecular flow to the hydrodynamic solution.
The physics to be presented depends on the ratio of time step $\Delta t$ over the particle collision time $\tau$.
But, the different choices of the initial condition $f_0$ at the beginning of each time step determines the different evolution mechanism,
i.e., the macroscopic flow solver, or a multiple scale evolution model.

For the linear advection-diffusion Eq.(\ref{advection-diffusion}),
the corresponding scheme used for its solution is the gas-kinetic scheme (GKS) for the update of macroscopic flow variable \cite{xu2001}, which is basically a NS solver
in the continuum flow regime.
Here, the initial condition $f_0$ in the solution Eq.\eqref{solution} is constructed based on the Chapman-Enskog
expansion, such as Eq.(\ref{CE}). This assumption automatically projects the distribution function to the fluid element modeling,
where a small deviation from the
equilibriums state is used for the capturing of diffusive effect.
 For the direct modeling UGKS, the initial condition $f_0$ is known through the update of the gas distribution function in Eq.(\ref{Finite-volume-f}).
Therefore, the departure from the equilibrium depends on the scale, such as the ratio of $\Delta t/\tau$. Note that the cell size and time step are the modeling scales of UGKS.

The GKS is a direct macroscopic  Eq.(\ref{advection-diffusion}) solver through the update of Eq.(\ref{Finite-volume-u}) alone.
The interface flux is based on solution Eq.\eqref{solution} with the adoption of the Chapman-Enskog expansion for its initial condition $f_0$.
The use of the Chapman-Enskog expansion makes GKS solve the advection-diffusion equation only.
For the UGKS, there is no such an assumption about the form of the initial gas distribution function,
and the real scale-dependent  distribution function is followed  for its evolution.
The capability of capturing multiscale physics from UGKS is mainly due to the scale dependent evolution solution Eq.(\ref{solution})
for the cell interface flux evaluation, which depends on the ratio of the time step $\Delta t$ over the particle collision time $\tau$.
The capturing of different physics can be
easily understood from the solution Eq.(\ref{solution}).
In the kinetic regime, i.e., $\Delta t \leq \tau$, the particle free transport from$f_0$ in
Eq.(\ref{solution}) contributes mainly for the flux function.
For the scale with $\Delta t \geq \tau$, the collision will gradually take effect. In the hydrodynamic limit, i.e.,
$\Delta t \gg \tau$, the NS gas distribution from the equilibrium state integration in Eq.(\ref{solution}) will play a dominant role.
Therefore, the solution provided in Eq.(\ref{solution}) depends on the
ratio $\Delta t /\tau$ or $\Delta x / l$. In other words, with the cell size $\Delta x \gg l$ and time step $\Delta t \gg \tau$, the multiple particle
collision is included in the integral solution Eq.(\ref{solution}), which is beyond the binary collision model in the full Boltzmann collision term.
In the transition regime with a cell resolution of multiple mean free path,
the solution in Eq.(\ref{solution}) includes the effect of multiple collisions for the individual particle.
In terms of the flow modeling, the UGKS presents a direct modeling equation in all scales,
which is beyond the scale for the derivation of the Boltzmann equation.
Even though the GKS has the same evolution mechanism from the particle free transport to the hydrodynamic evolution, the use of the Chapman-Enskog expansion confines its applicable regime to the
near equilibrium only in the macroscopic scale.
With a different approximation for the initial gas distribution function $f_0$,
the UGKS can give a valid solution in all regimes and use a time step with a fixed CFL number, which is independent of the Reynolds number.
However, for the GKS, same as other explicit NS  solver, its solution is limited to the continuum flow regime and the time step is severely constrained at the low Reynolds number case.

\section{Dynamical differences in case studies}

\subsection{Linear advection-diffusion process and the corresponding multiple scale solution}

%\begin{figure}[!htb]
%\centering
%\includegraphics[bb=47 79 610 522, clip,width=10cm]{space-evolution.eps}
%\caption{Multiscale flow evolution in space with different resolution.}
%\label{fig:space-evolution}
%\end{figure}

To study the dynamical differences quantitatively from different modeling, we first solve Eq.\eqref{advection-diffusion} for the advection diffusion solution
in the domain $x\in[-1,3]$ with the periodic boundary condition. The initial condition is set as
\begin{equation}\label{initial}
  u_0(x)=4+\frac{8}{\pi}\sin\left(\frac{\pi}{2}x\right)+\frac{16}{3\pi}\sin\left(3\frac{\pi}{2}x\right).
\end{equation}
For the linear advection-diffusion equation, the analytic solution is given by
\begin{equation}\label{analytic}
  u(x,t)=4+\frac{8}{\pi}\mathrm{e}^{-\frac{\pi^2}{4}\nu t}\sin\left(\frac{\pi}{2}(x-at)\right)+\frac{16}{3\pi}\mathrm{e}^{-\frac{9\pi^2}{4}\nu t}\sin\left(3\frac{\pi}{2}(x-at)\right).
\end{equation}

\begin{figure}
\centering
\includegraphics[width=0.35\textwidth]{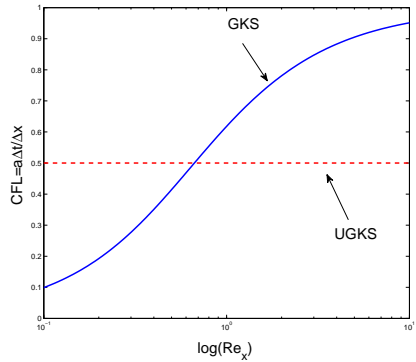}
\caption{Maximum CFL number for GKS (NS) and UGKS with different Reynolds numbers.}
\label{stability}
\end{figure}

Based on the GKS (advection-diffusion solution) and the UGKS (multiscale modeling solution),
in the high Reynolds number limit, both results
are identical in the hydrodynamic regime. The current study is mostly on the transition regime at the low Reynolds number limit, where both
fluid element and particle penetration play important role.
In the low Reynolds number regime, the stability condition for GKS, like many other advection-diffusion solvers,  becomes restricted.
The time step is limited by
$\Delta t < (\Delta x)^2/(2\nu) $. However, for the the UGKS, a necessary stability condition is due to the particle velocity range used to discretize the velocity space.
In other words, for the UGKS, the time step is determined by the CFL condition only,
\begin{equation}\label{stability-ugks}
  \Delta t \le \frac{\Delta x}{|a|+3\sqrt{\theta}}.
\end{equation}
The stability condition for GKS and UGKS is shown in Fig. \ref{stability} with $a=3$ and $\theta =1.0$ under different cell's Reynolds number, i.e., $Re_{x} = a \Delta x /(\theta \tau) $.

\begin{figure}
\centering
\includegraphics[width=0.3\textwidth]{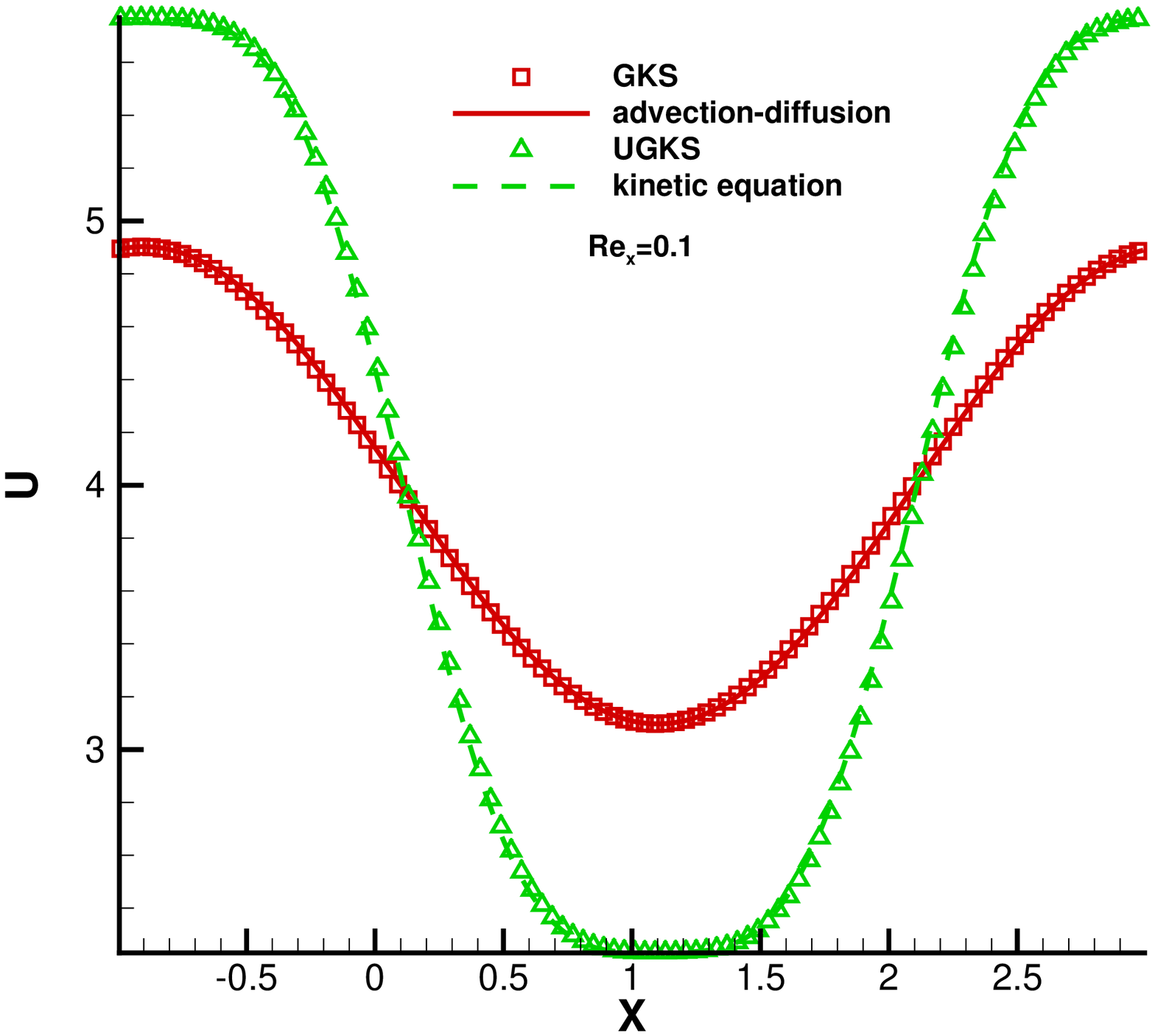}
\includegraphics[width=0.3\textwidth]{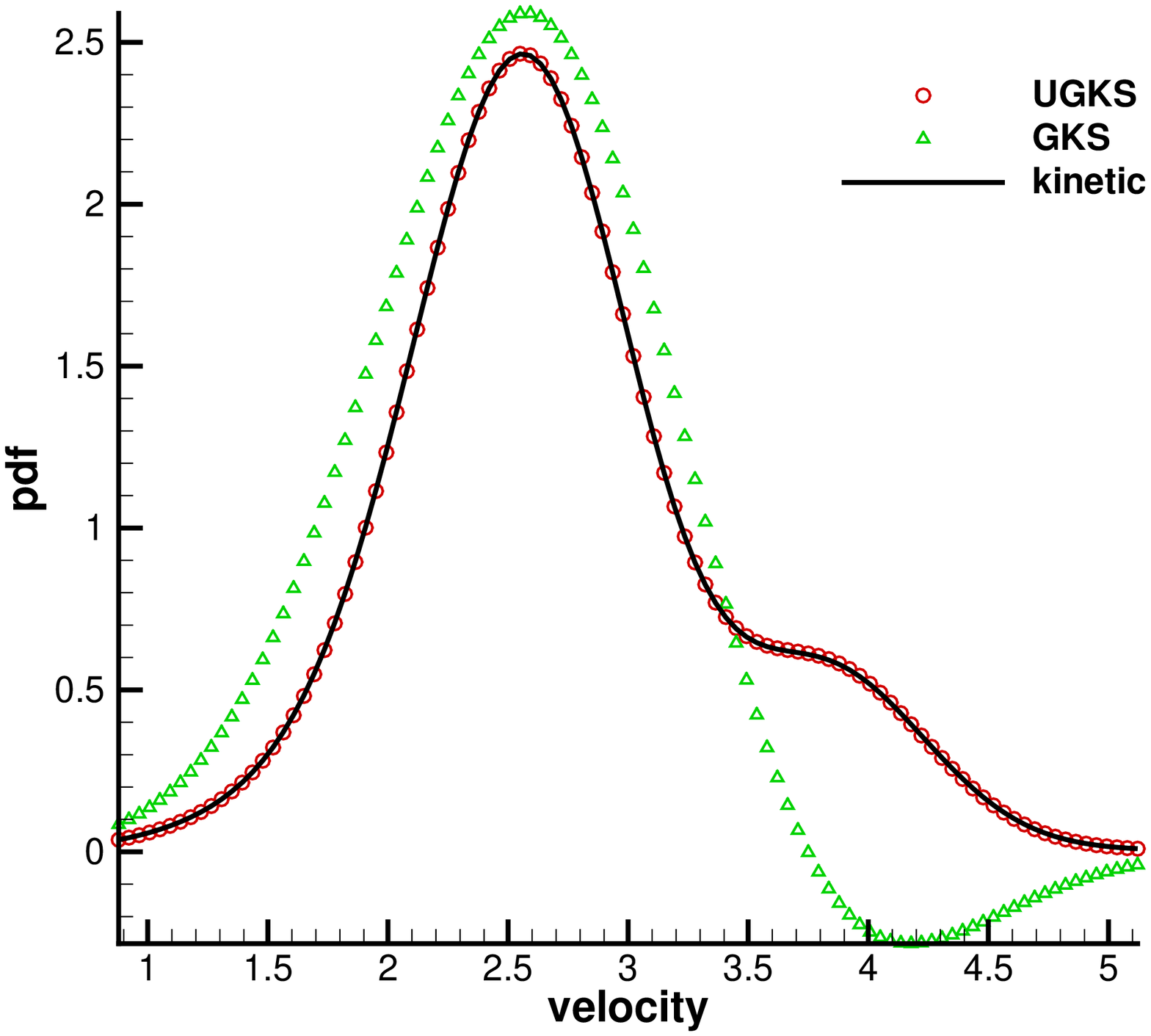}\\
\includegraphics[width=0.3\textwidth]{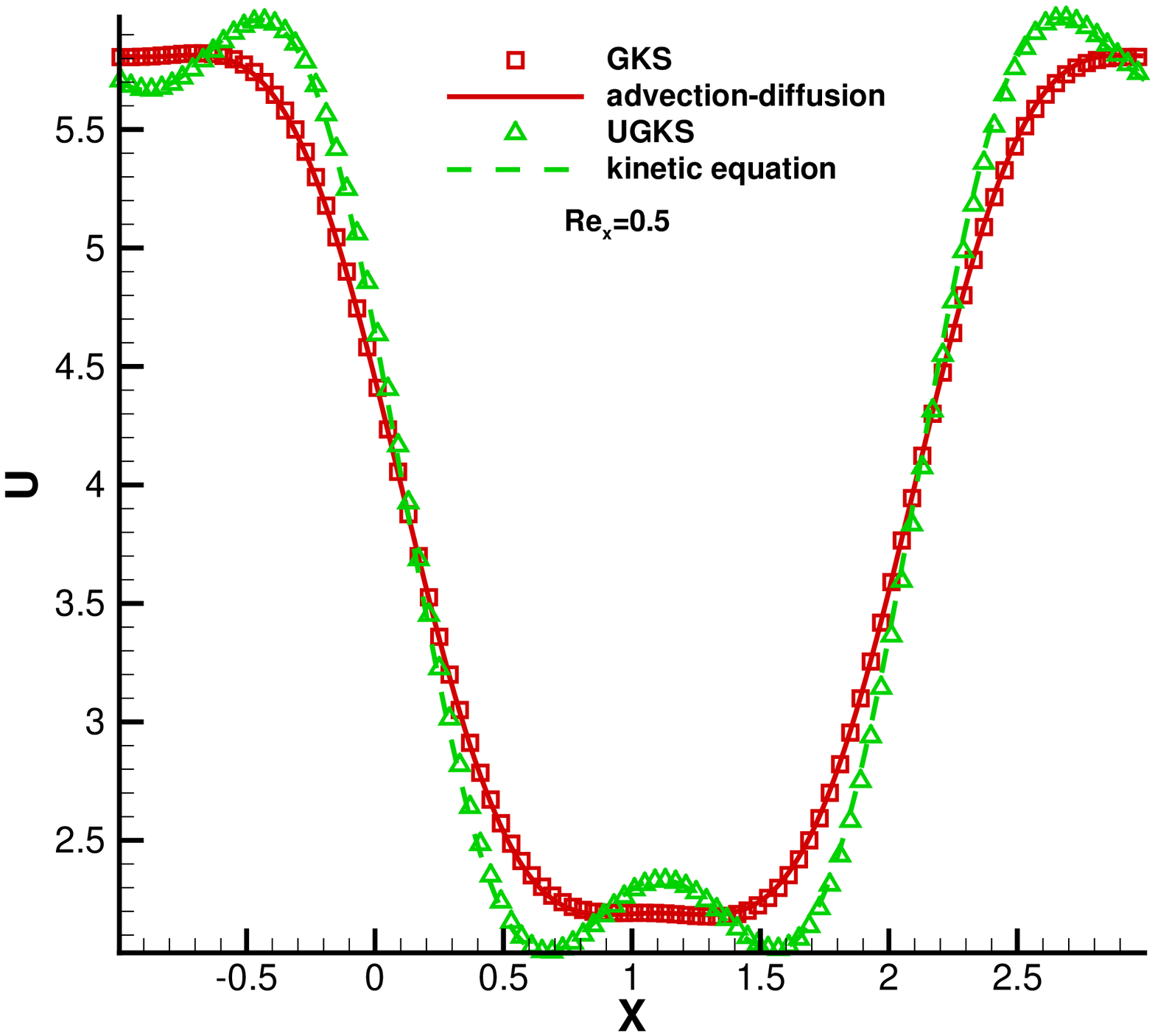}
\includegraphics[width=0.3\textwidth]{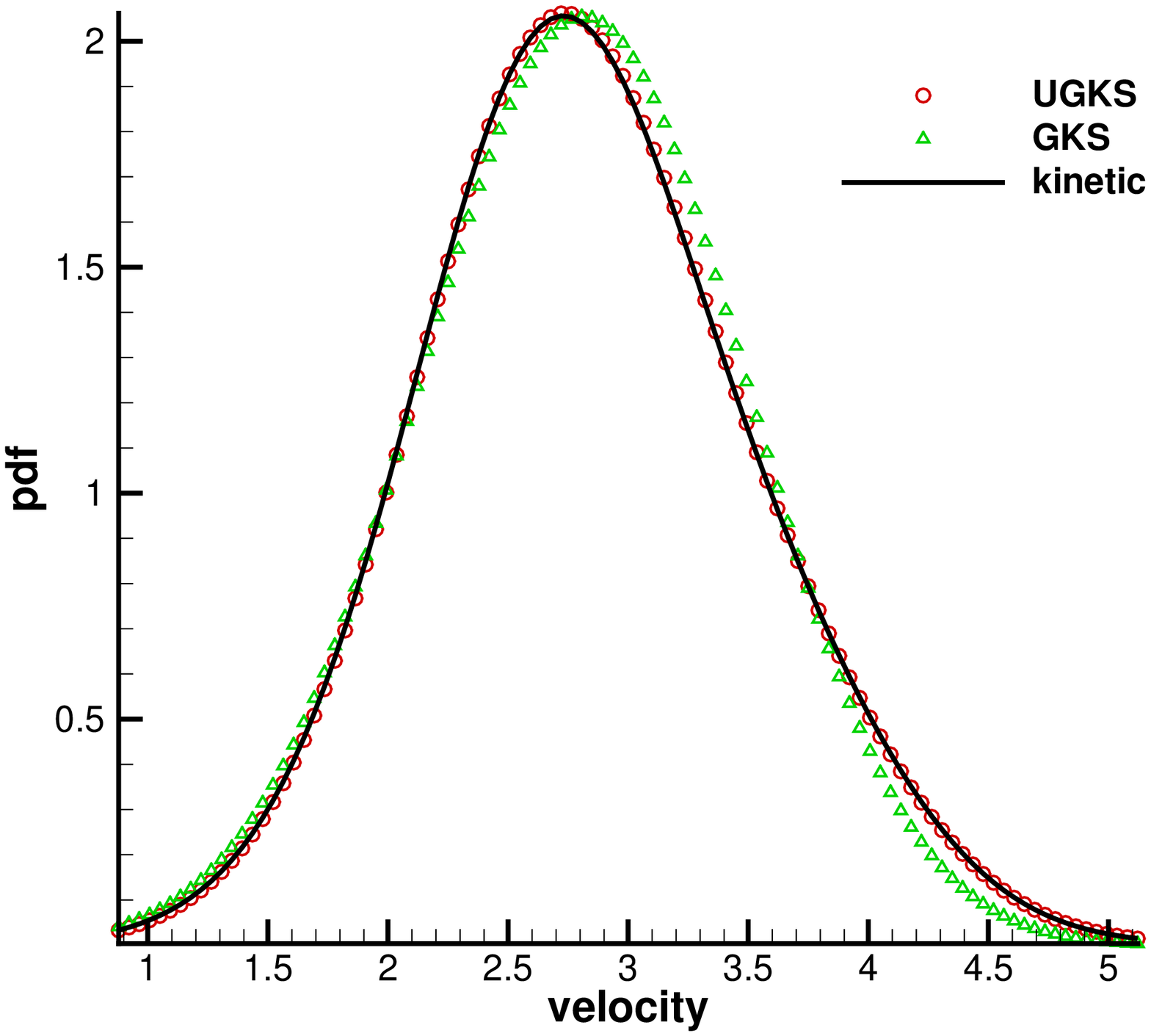}\\
\includegraphics[width=0.3\textwidth]{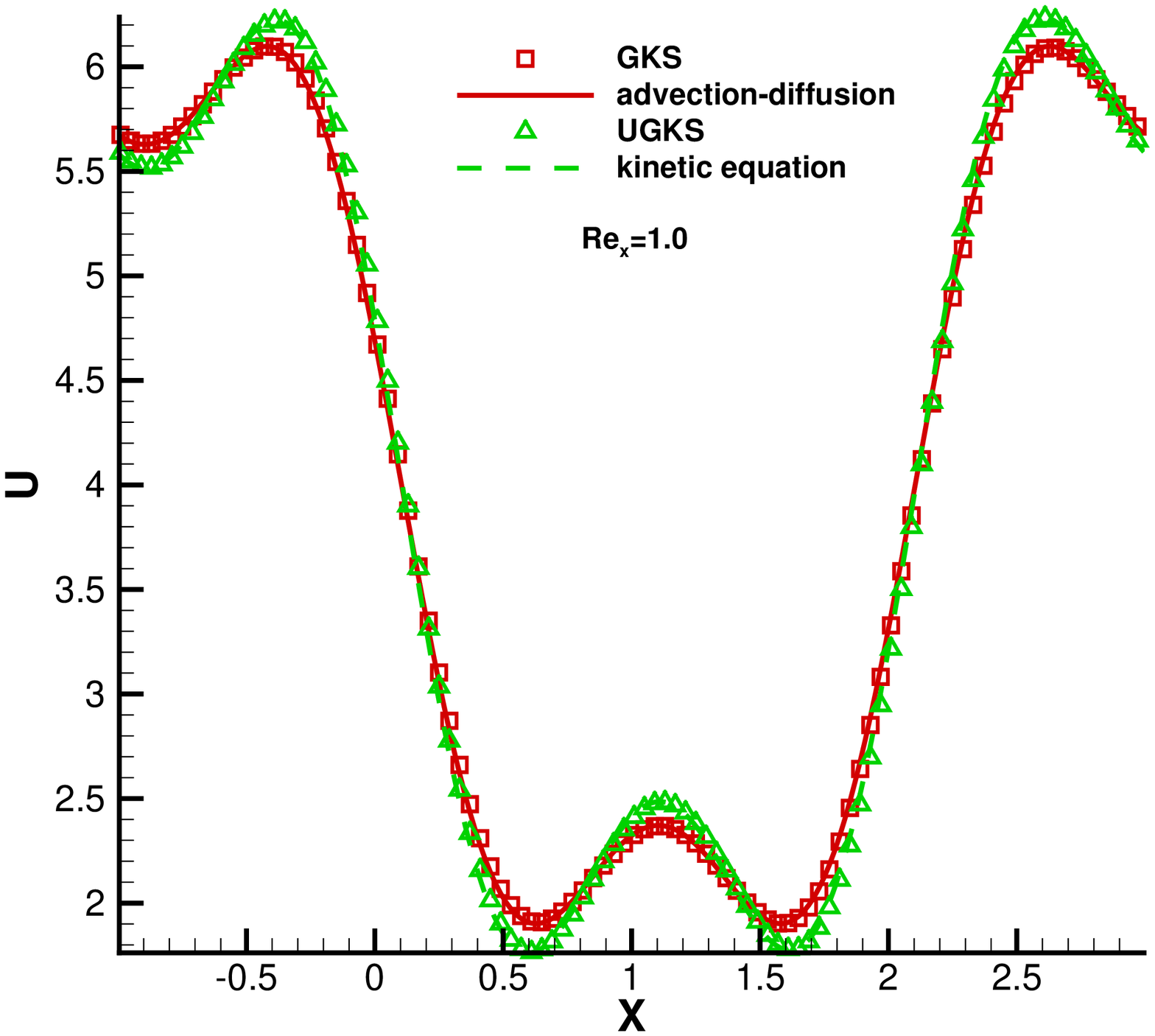}
\includegraphics[width=0.3\textwidth]{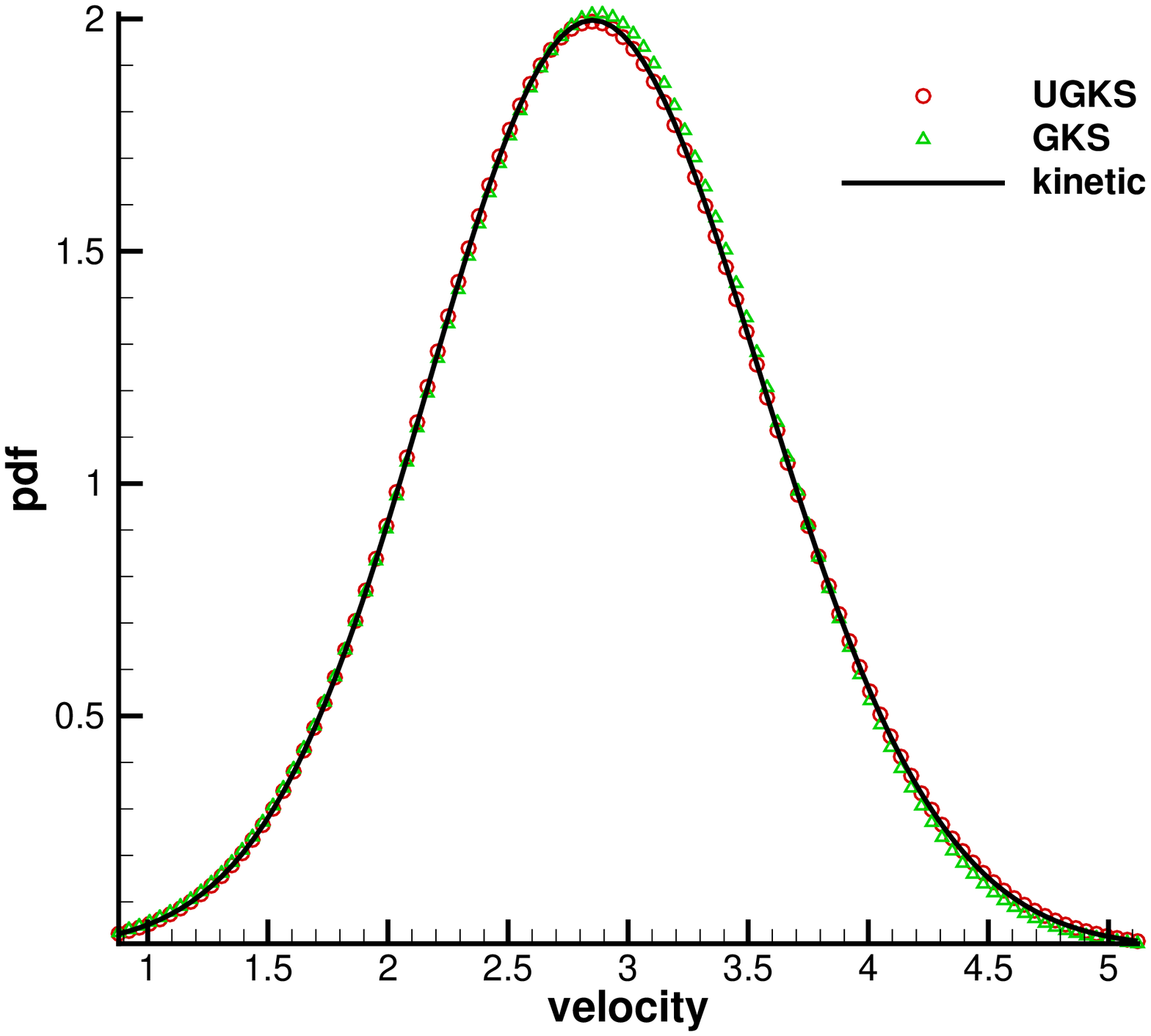}\\
\includegraphics[width=0.3\textwidth]{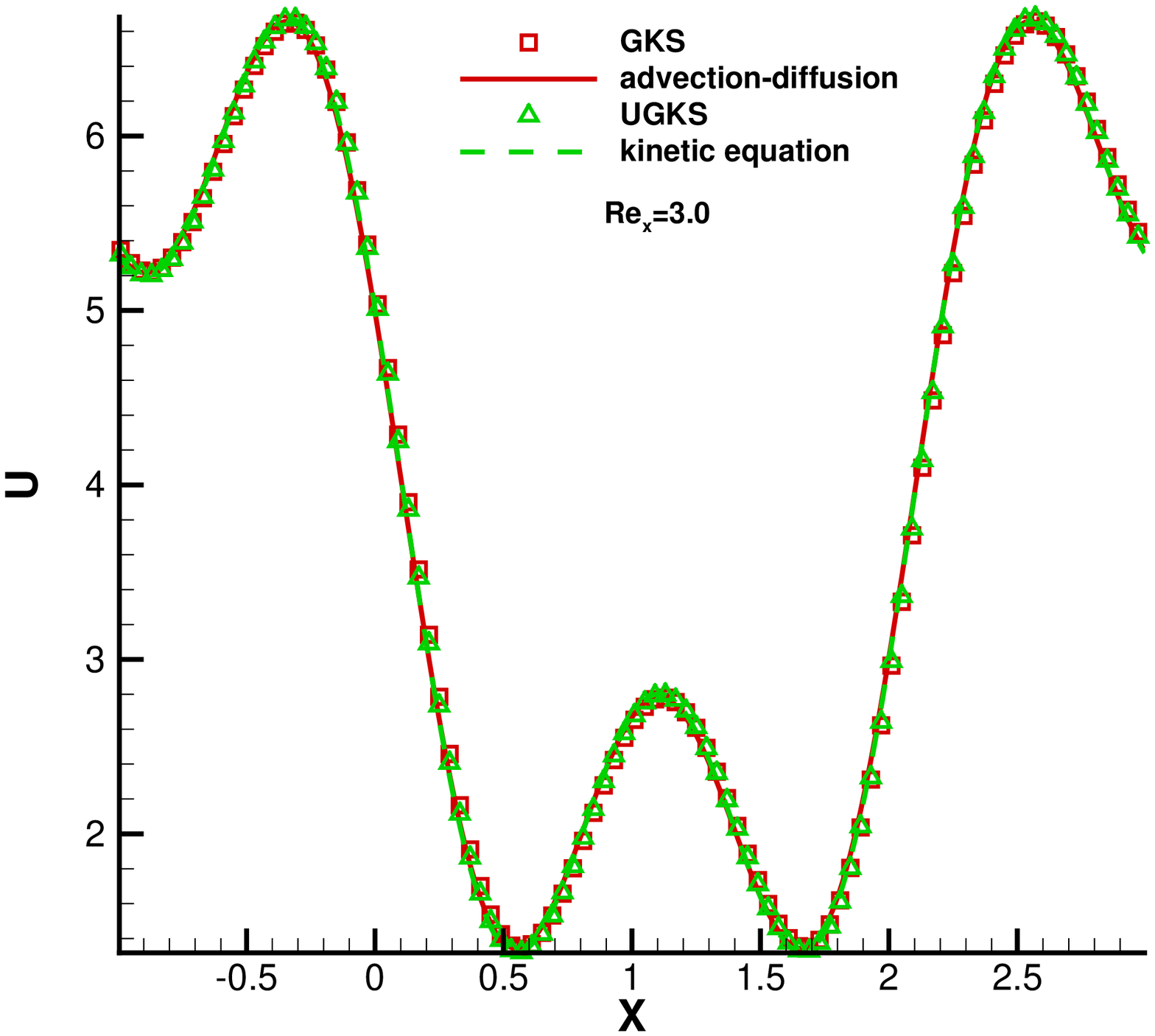}
\includegraphics[width=0.3\textwidth]{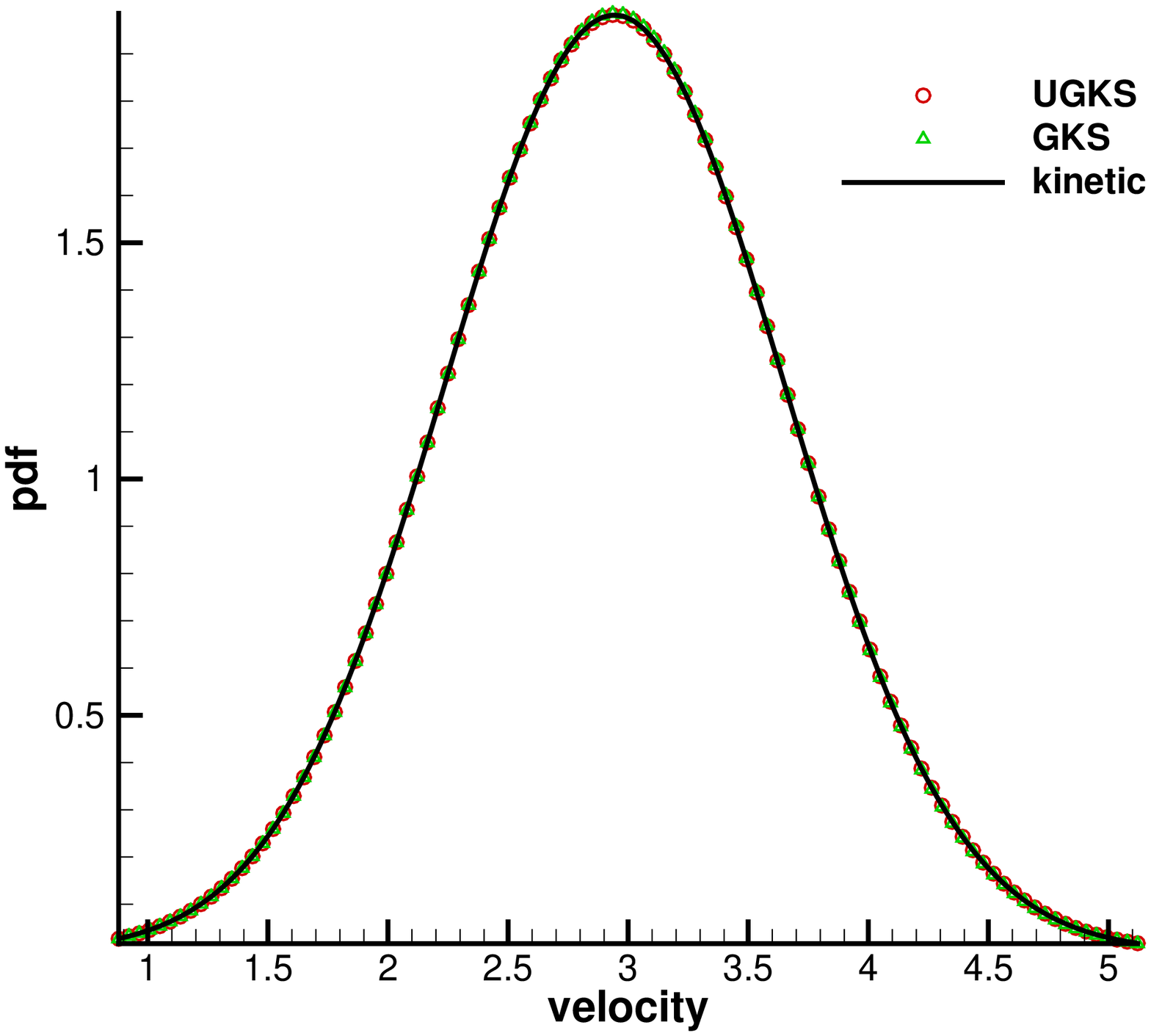}
\caption{Left column shows the comparison of the macroscopic quantity $u$, symbols are the numerical solutions, lines are the exact solutions.
Right column shows the comparison of the velocity distribution function at $x=2$, lines are the exact solutions of kinetic equation.
From top to bottom, the corresponding cell Reynolds numbers are ${Re}_x = a \Delta x /(\theta \tau) = 0.1, 0.5, 1.0 $, and $3.0$.
}
\label{compare}
\end{figure}

Under the stability condition, the solutions from the above two schemes behave differently with the variation of Reynolds numbers.
By setting the parameters $a=3$, $\theta =1$, $\Delta x=0.04$, we compare the solutions of GKS and UGKS with cell Reynolds number $\text{Re}_x=0.1, 0.5, 1.0, 3.0$.
In Fig. \ref{compare}, we plot the macroscopic quantity $u$  and the velocity distributions at $t=0.7$ and $x=2$.
At high Reynolds number regime, both advection-diffusion equation and UGKS solutions are consistent, and the macroscopic description is a valid model.
When Reynolds number decreases to be less than $1$, the advection-diffusion solution deviate from the UGKS solution.
Especially, when $\text{Re}_x=0.1$, the distribution function corresponding to the advection diffusion model, which is obtained from the Chapman-Enskog expansion,
can become negative in certain particle velocity region.
The possible negative particle velocity distribution and the severe time step limitation indicate that
the advection-diffusion equation is not applicable  for capturing flow physics in this regime.
For low Reynolds number flow, the time step used in UGKS is independent of Reynolds number, which is consistent with the physical reality.

\subsection{The NS solution and the multiscale  modeling solution}
\begin{figure}[!htb]
\centering
\includegraphics[width=8cm]{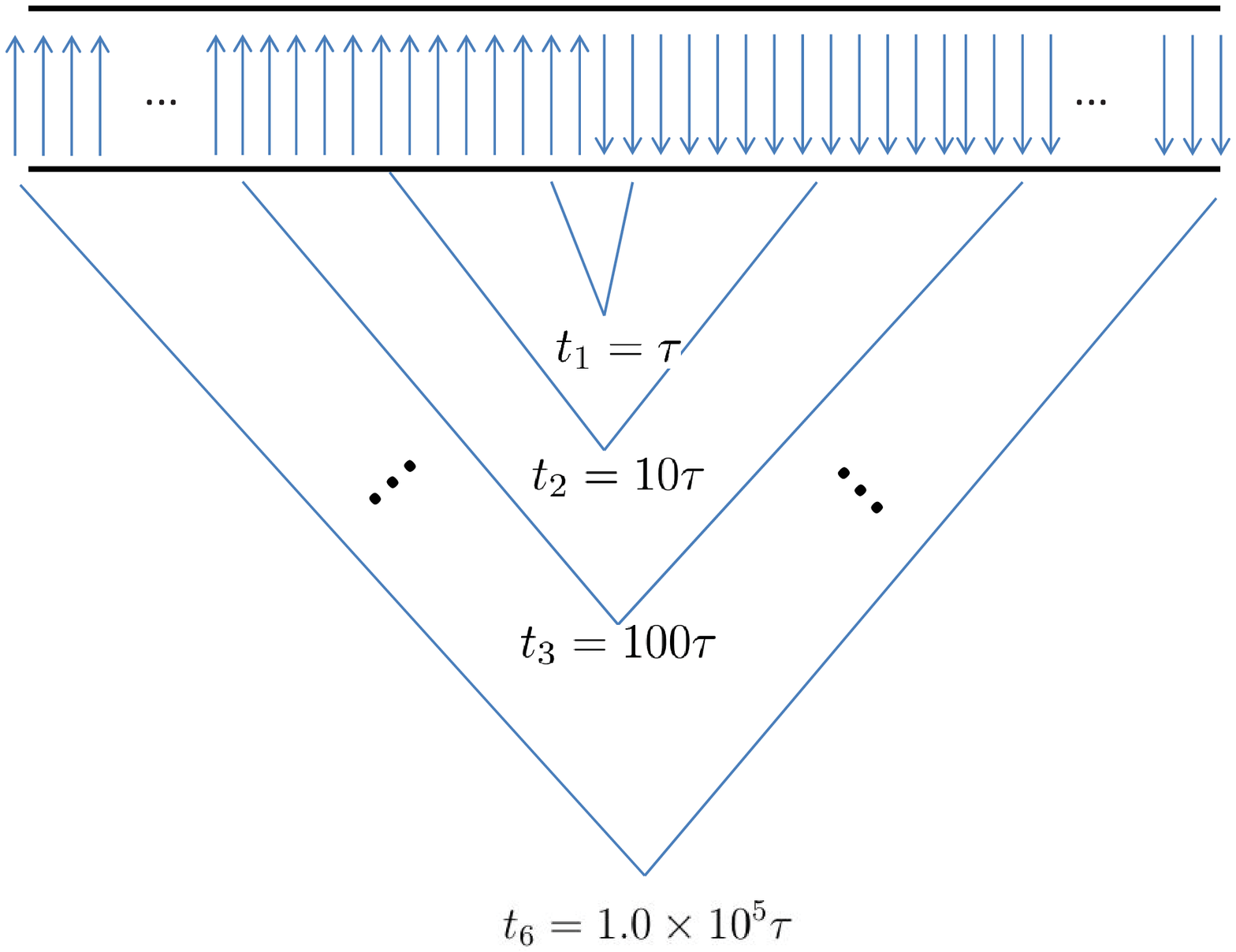}
\caption{Multiscale flow evolution for a shear layer, where $t_i$ is the evolution time and $\tau$ is the particle collision time.
The corresponding computational domain is changing with $t_i$ with different cell size $\Delta x$ relative to the particle mean free path $l_{mfp}$.}
\label{fig:time-evolution}
\end{figure}

We present the simulation results of a shear layer by GKS (NS)  and UGKS.
The GKS gives the Navier-Stokes solutions and the UGKS captures a multiscale physical solution.
The VHS model of argon gas is used in the simulation, and the Knudsen number $Kn = l/L$ is fixed to be $5.0\times 10^{-3}$.
The initial condition is given as
\begin{equation}\label{initial-condition}
  (\rho,U,V,T)=
  \begin{cases}
  (1.0,0,1.0,1.0)\qquad &x\leq0,\\
  (1.0,0,-1.0,0.5)\qquad &x>0.
  \end{cases}
\end{equation}

\begin{figure}
  \centering
  \includegraphics[width=0.4\textwidth]{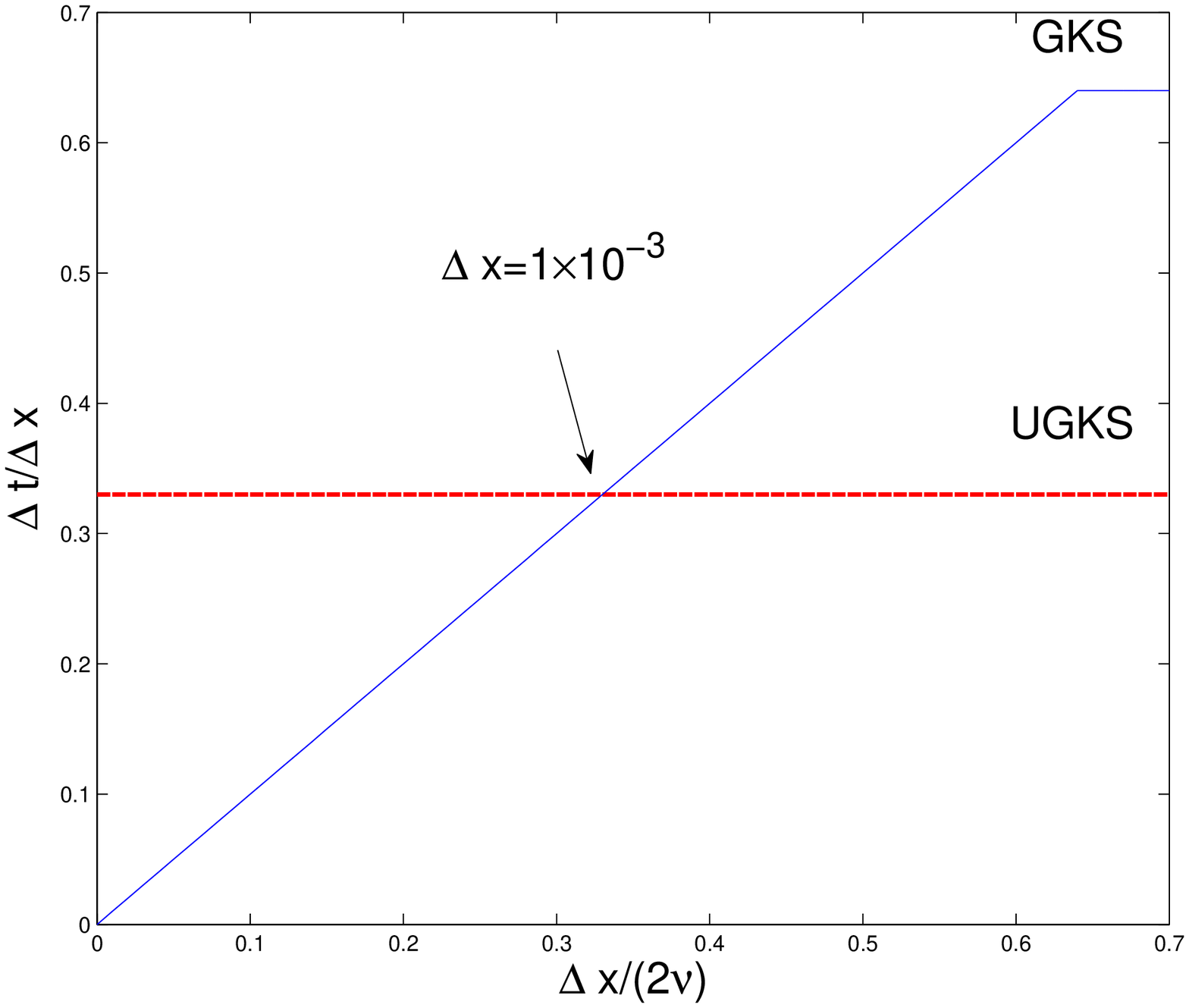}\\
  \caption{time step v.s. cell size for GKS (NS) and UGKS.}\label{fig:cfl}
\end{figure}

The mean free path is $l_{mfp}=5.0\times 10^{-3}$ and
the physical mean collision time is $\tau=3.36\times 10^{-3}$.
Since we are going to study a time-dependent multiple scale problem, the solutions to be obtained depend on
the output time, see Fig. \ref{fig:time-evolution}, where the computational domain with a fixed number of grid points is changing with the domain of influence from the initial singular point.
Same as the last case, the time steps used for the GKS and UGKS are shown in Fig. \ref{fig:cfl}, where the GKS or NS modeling has severe time step limitation
at low Reynolds number limit, but the UGKS uses a unform CFL number.

\begin{figure}
\centering
\includegraphics[width=0.4\textwidth]{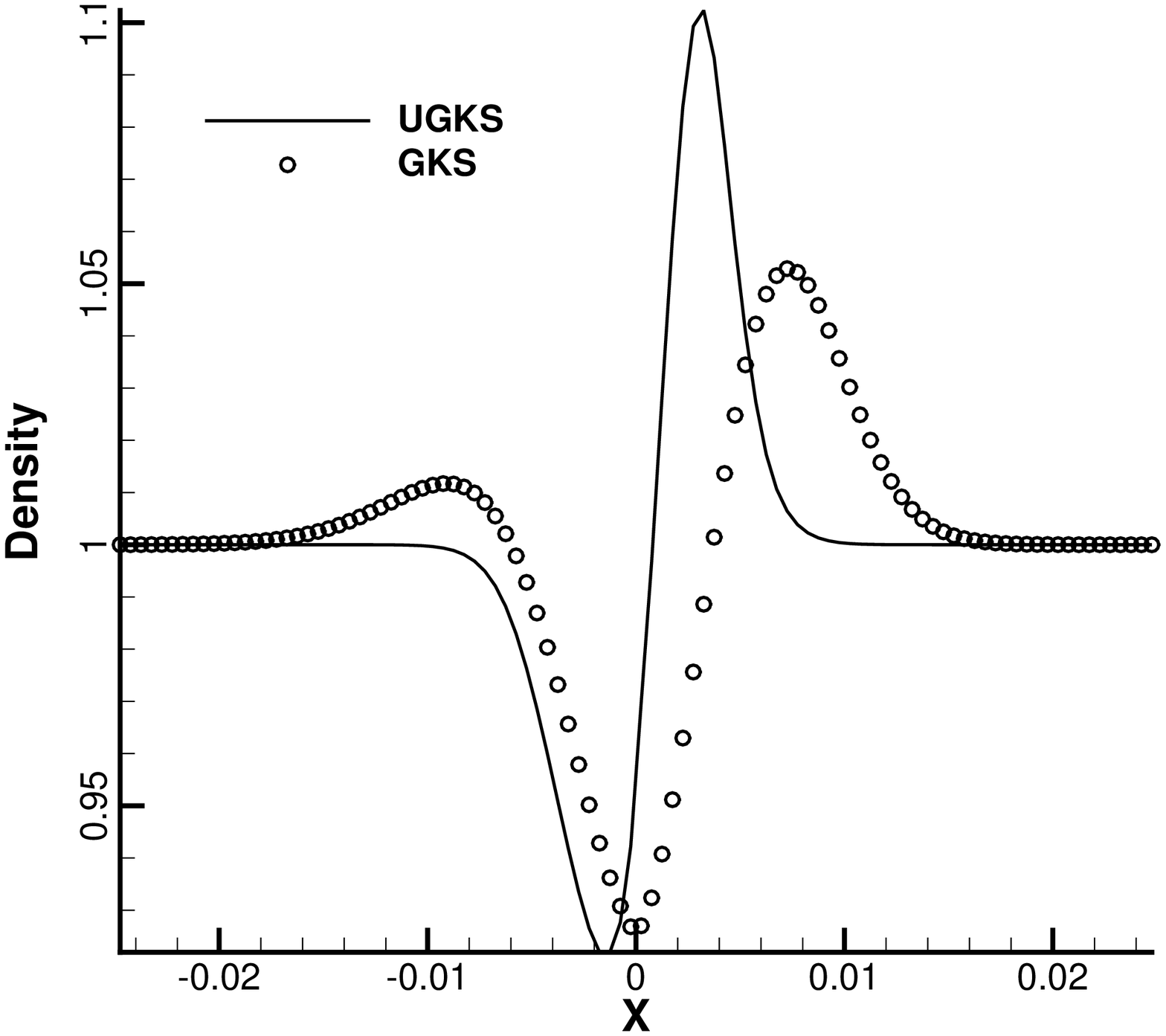}{a}
\includegraphics[width=0.4\textwidth]{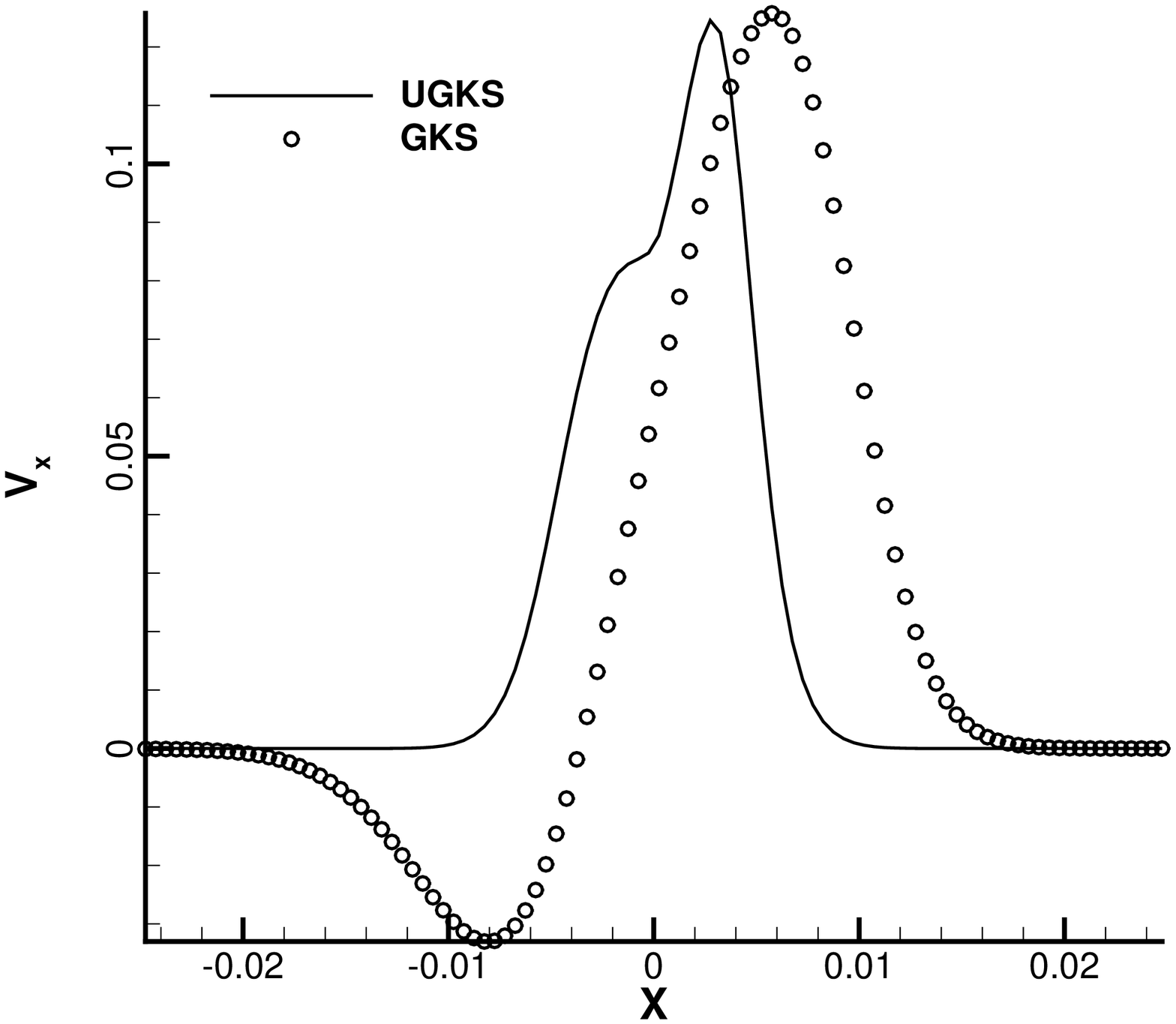}{b}\\
\includegraphics[width=0.4\textwidth]{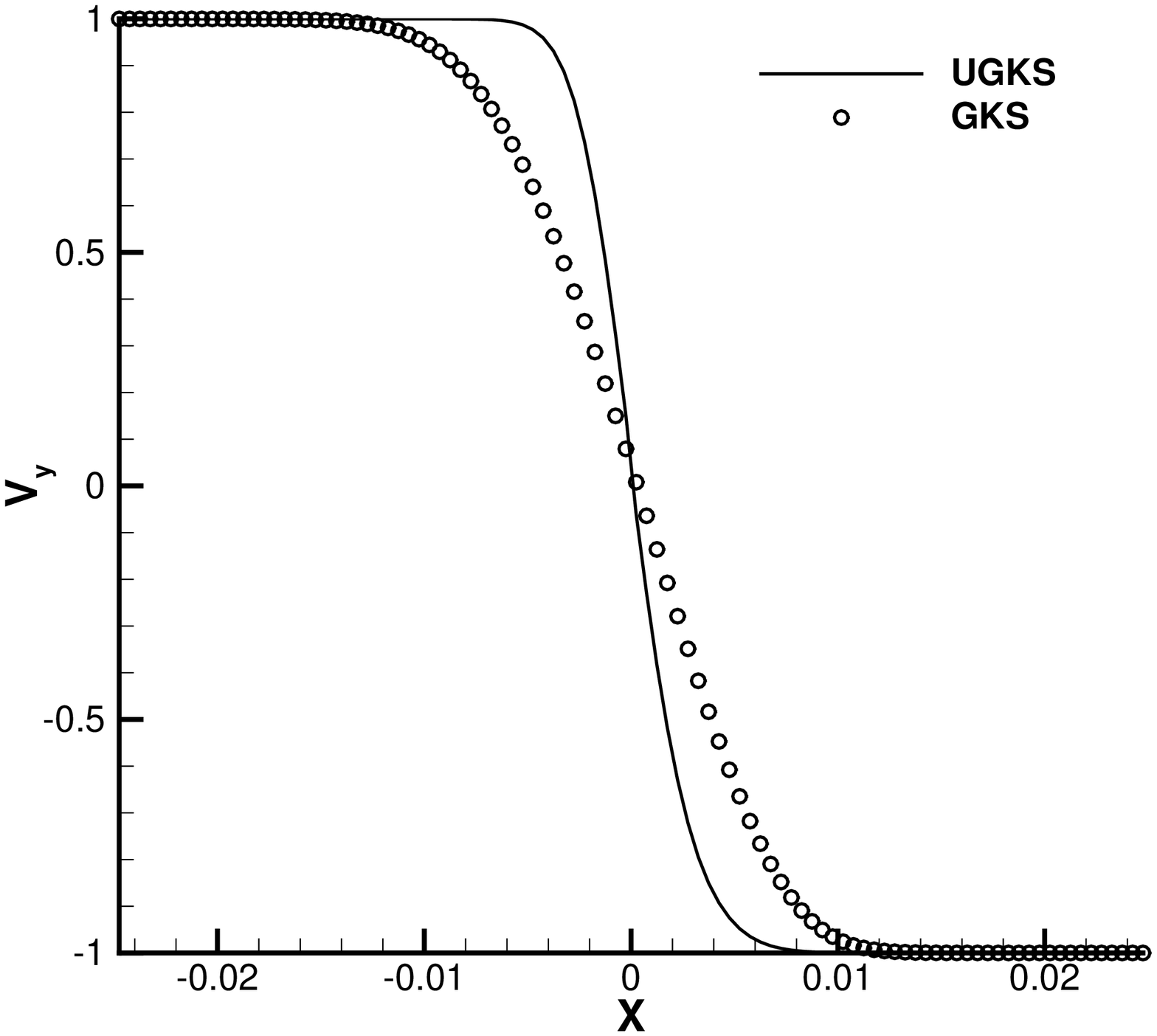}{c}
\includegraphics[width=0.4\textwidth]{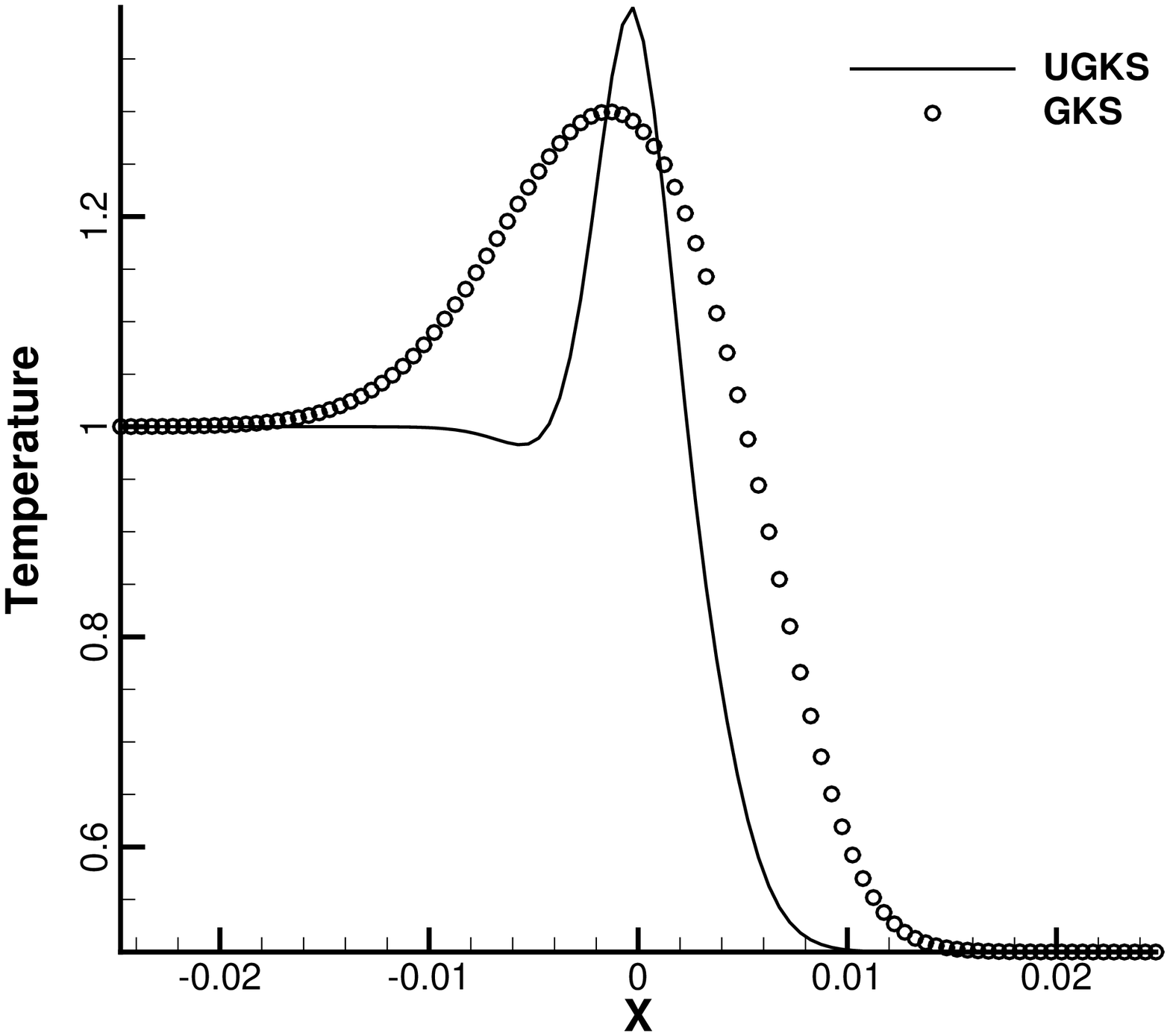}{d}\\
\includegraphics[width=0.4\textwidth]{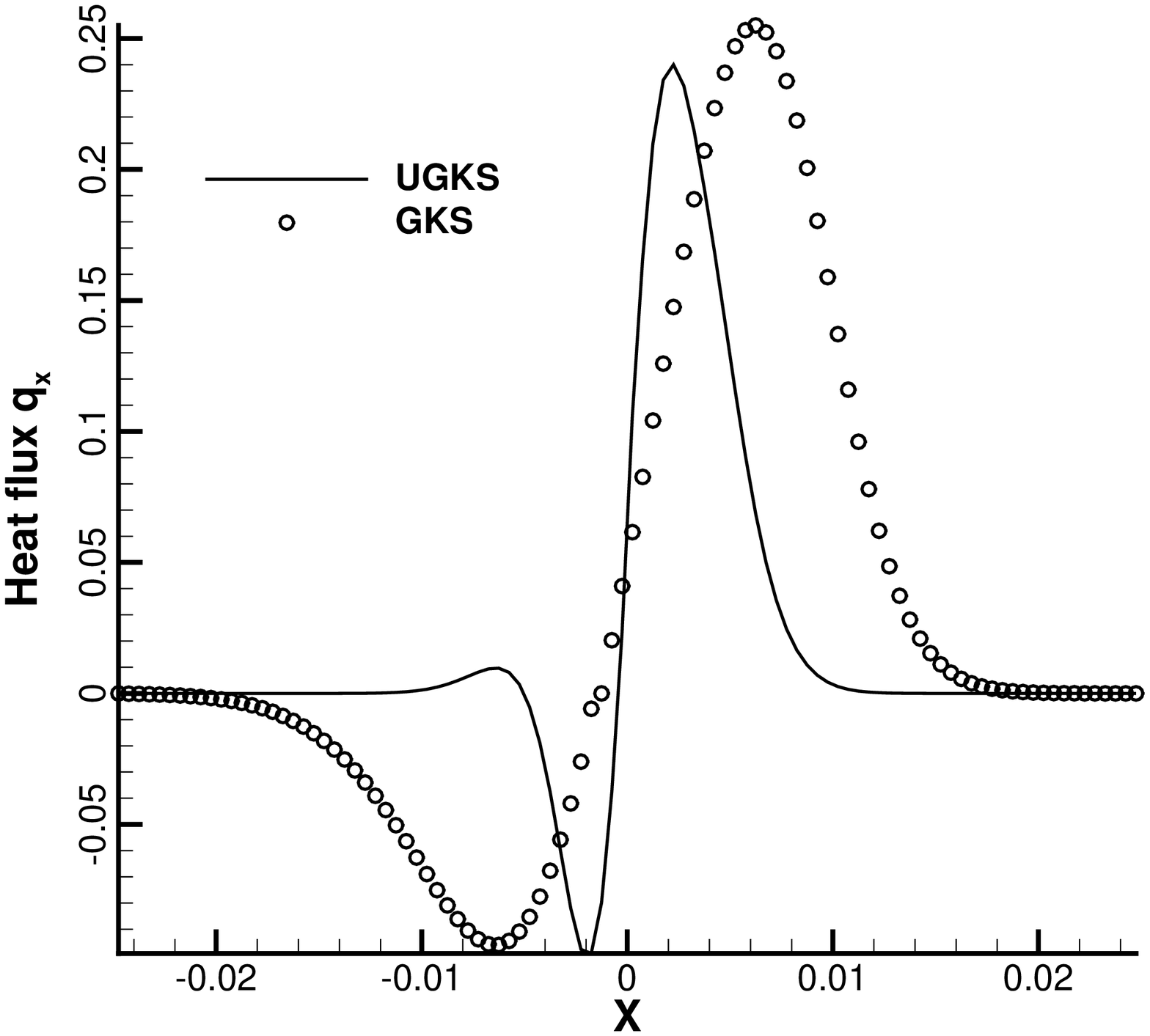}{e}
\includegraphics[width=0.4\textwidth]{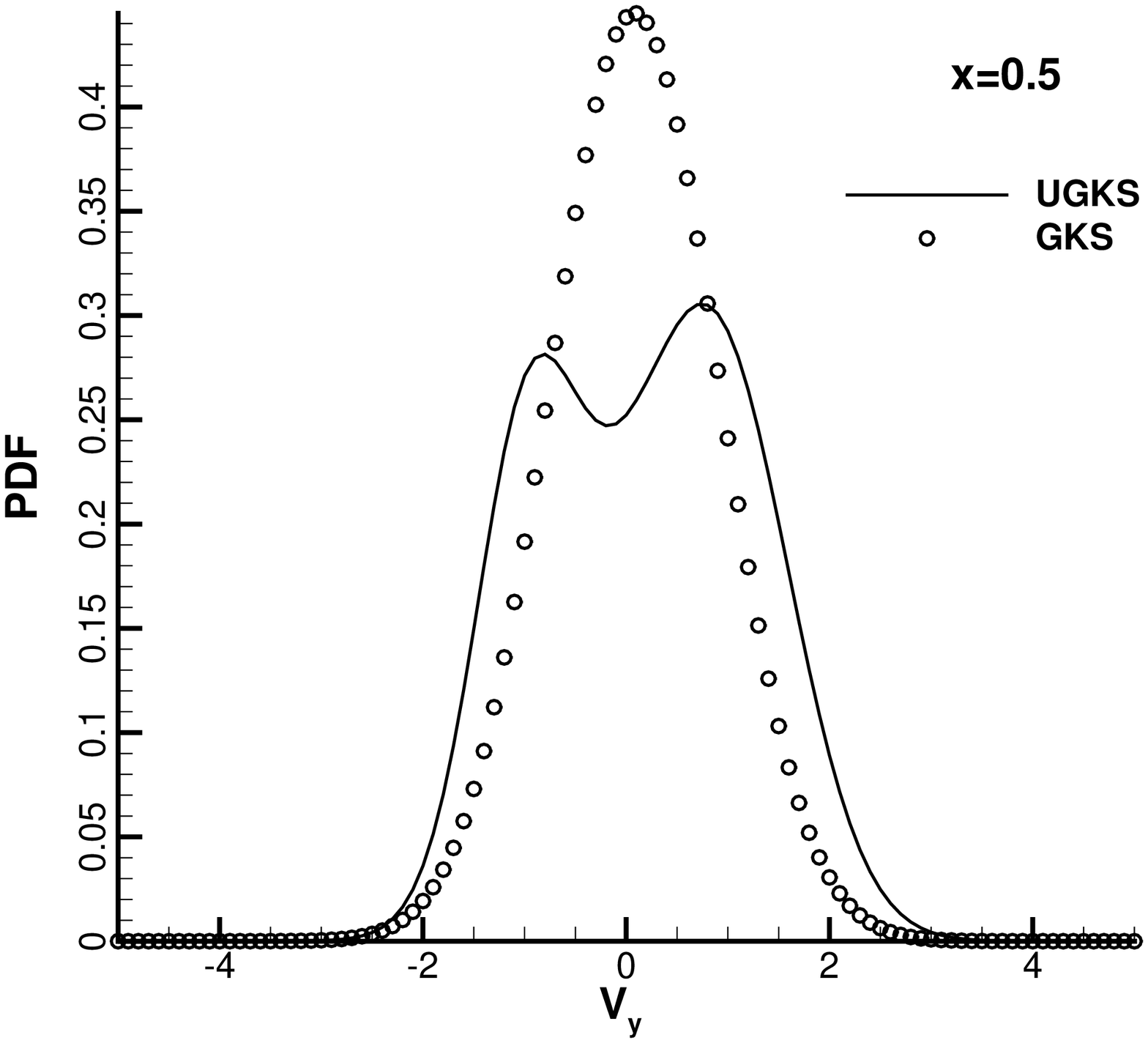}{f}
\caption{Results at $t=4\times 10^{-3}$ ($t/\tau=1.05$)
: a. density; b. x-velocity; c. y-velocity; d. temperature; e. x direction heat flux;
f. velocity distribution at $x=0.5$.
For GKS $\Delta x /l_{mfp}=0.1$, $\Delta t /\tau=1.1\times10^{-3}$,
and for UGKS $\Delta x/l_{mfp}=0.1$, $\Delta t/\tau=1\times 10^{-2}$.}\label{t3}
\end{figure}

\begin{figure}
\centering
\includegraphics[width=0.4\textwidth]{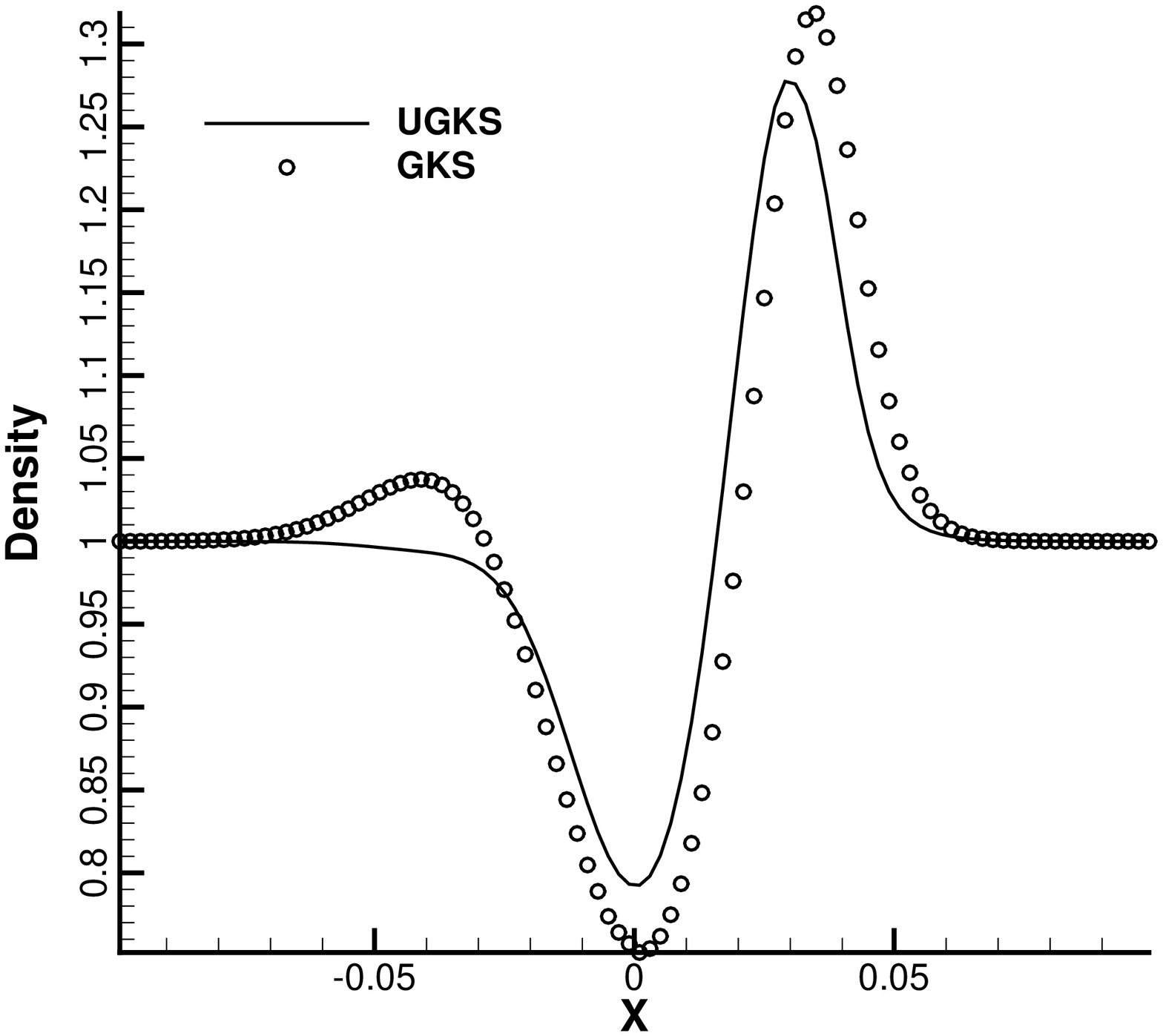}{a}
\includegraphics[width=0.4\textwidth]{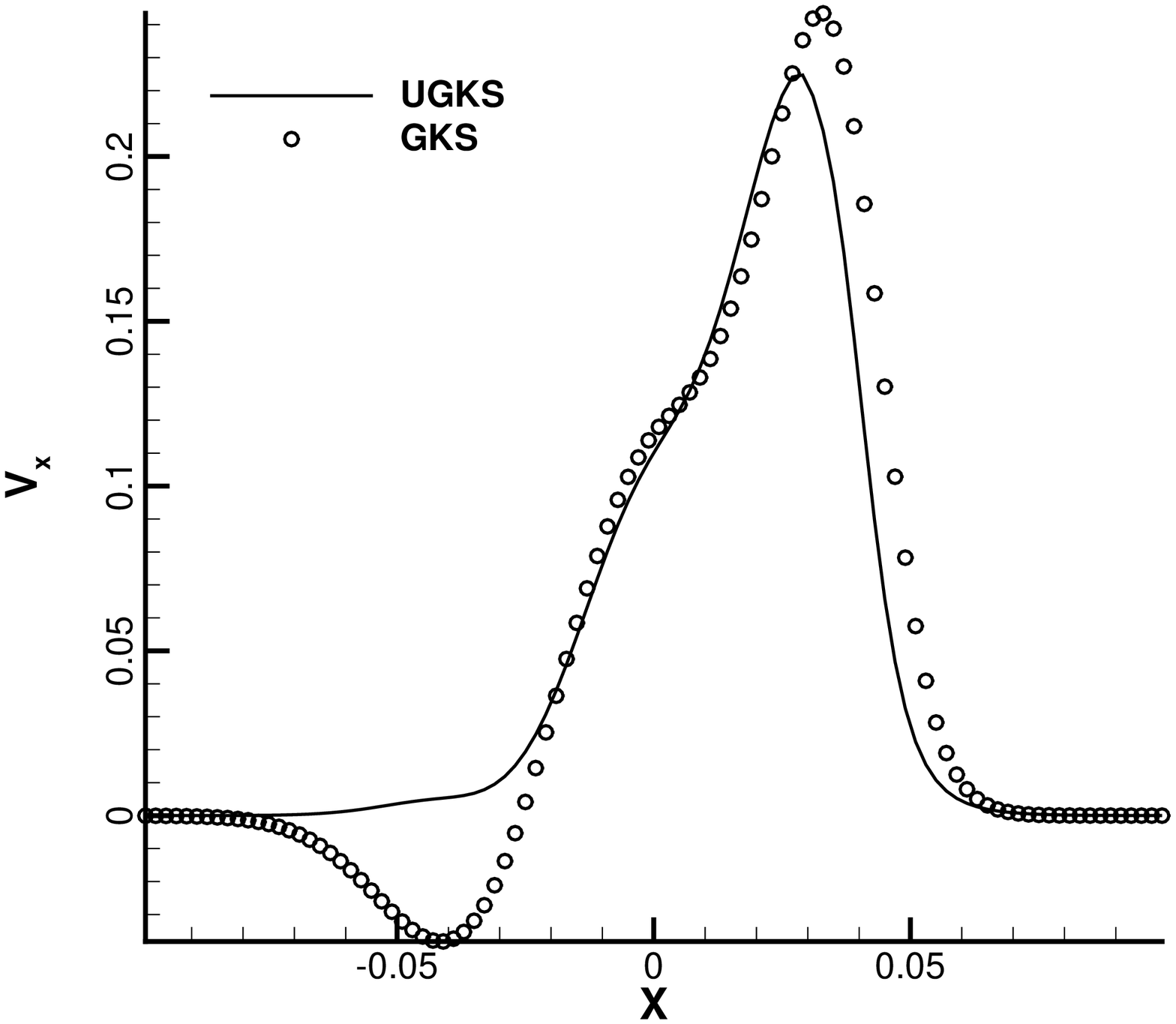}{b}\\
\includegraphics[width=0.4\textwidth]{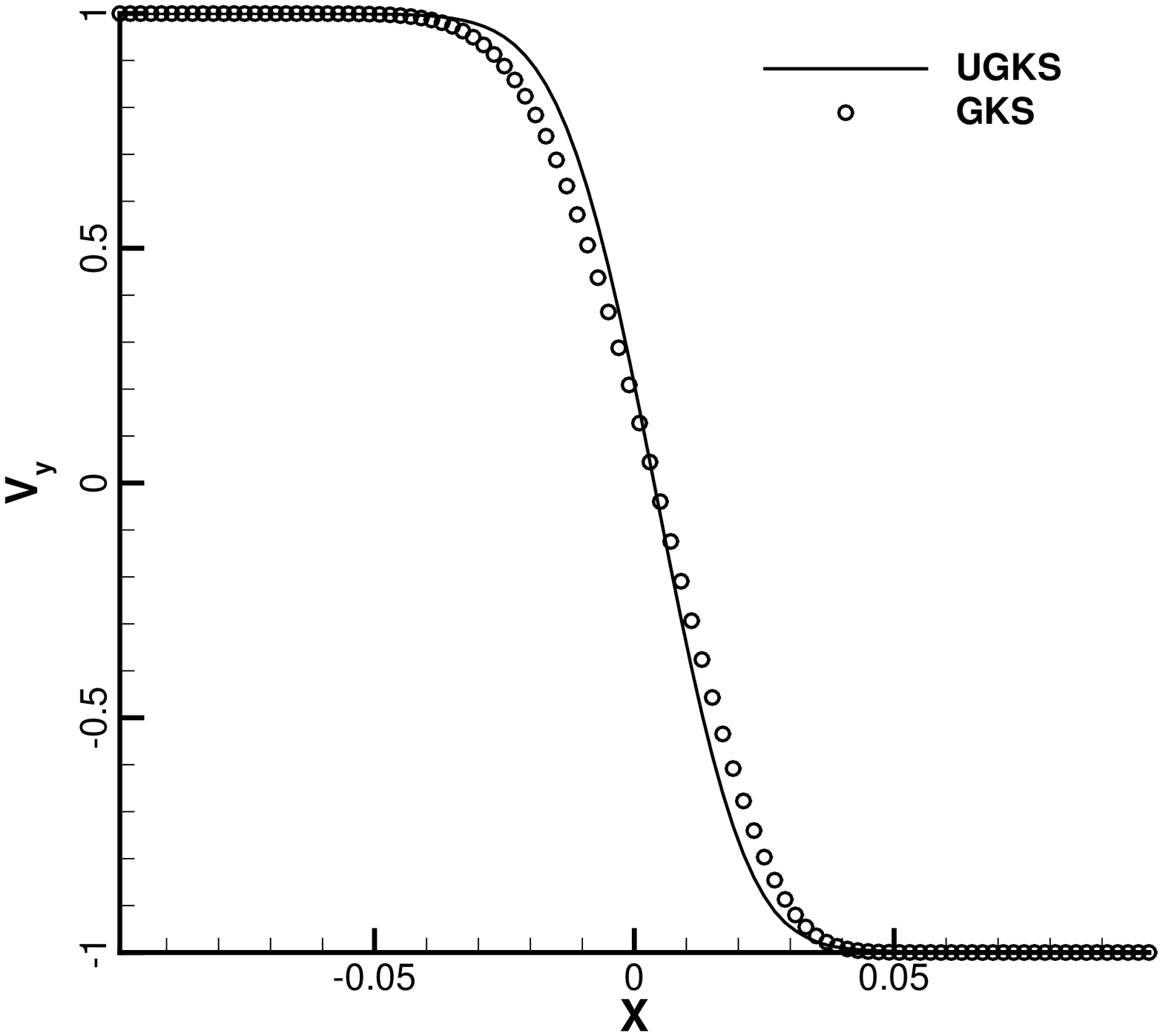}{c}
\includegraphics[width=0.4\textwidth]{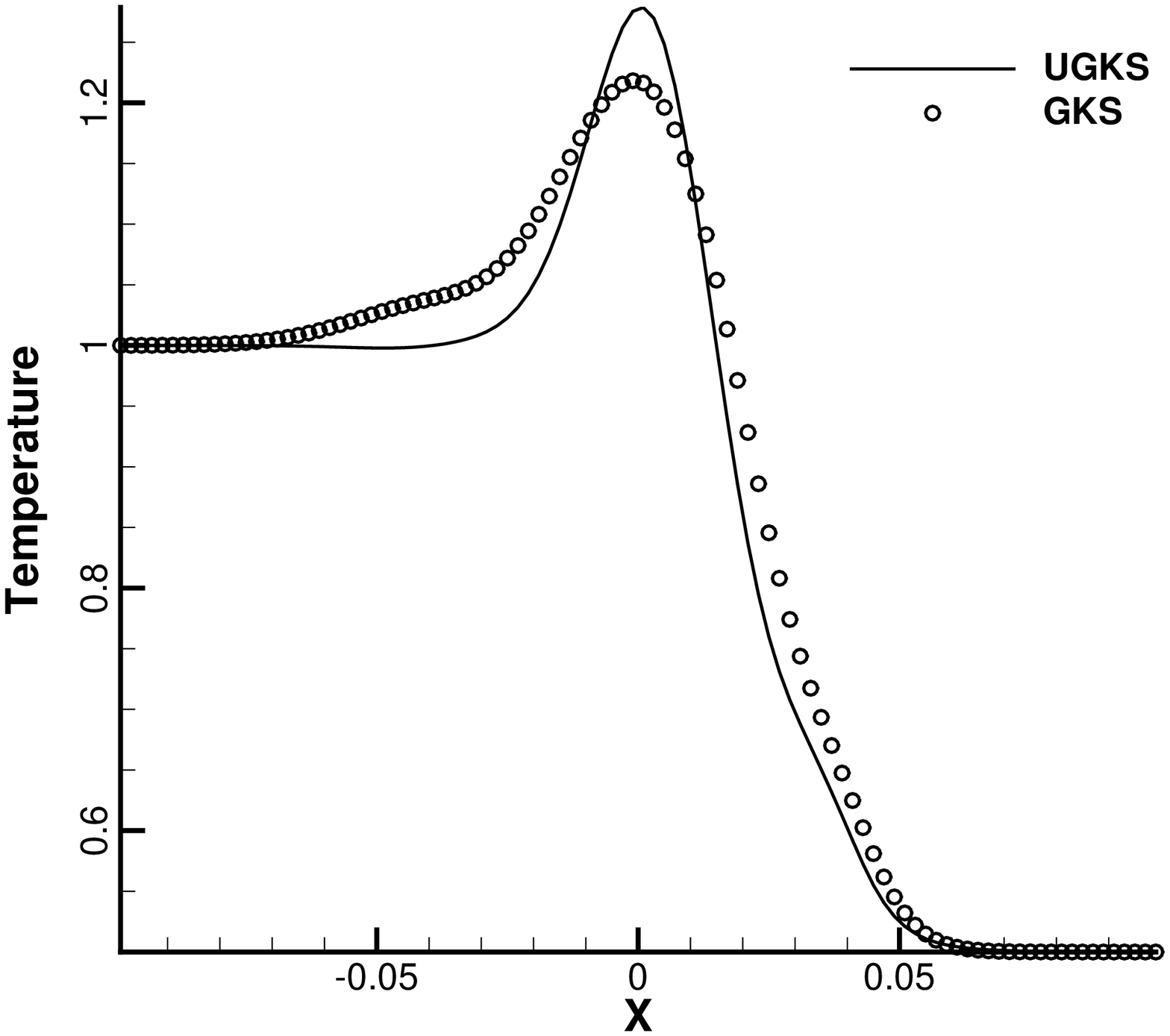}{d}\\
\includegraphics[width=0.4\textwidth]{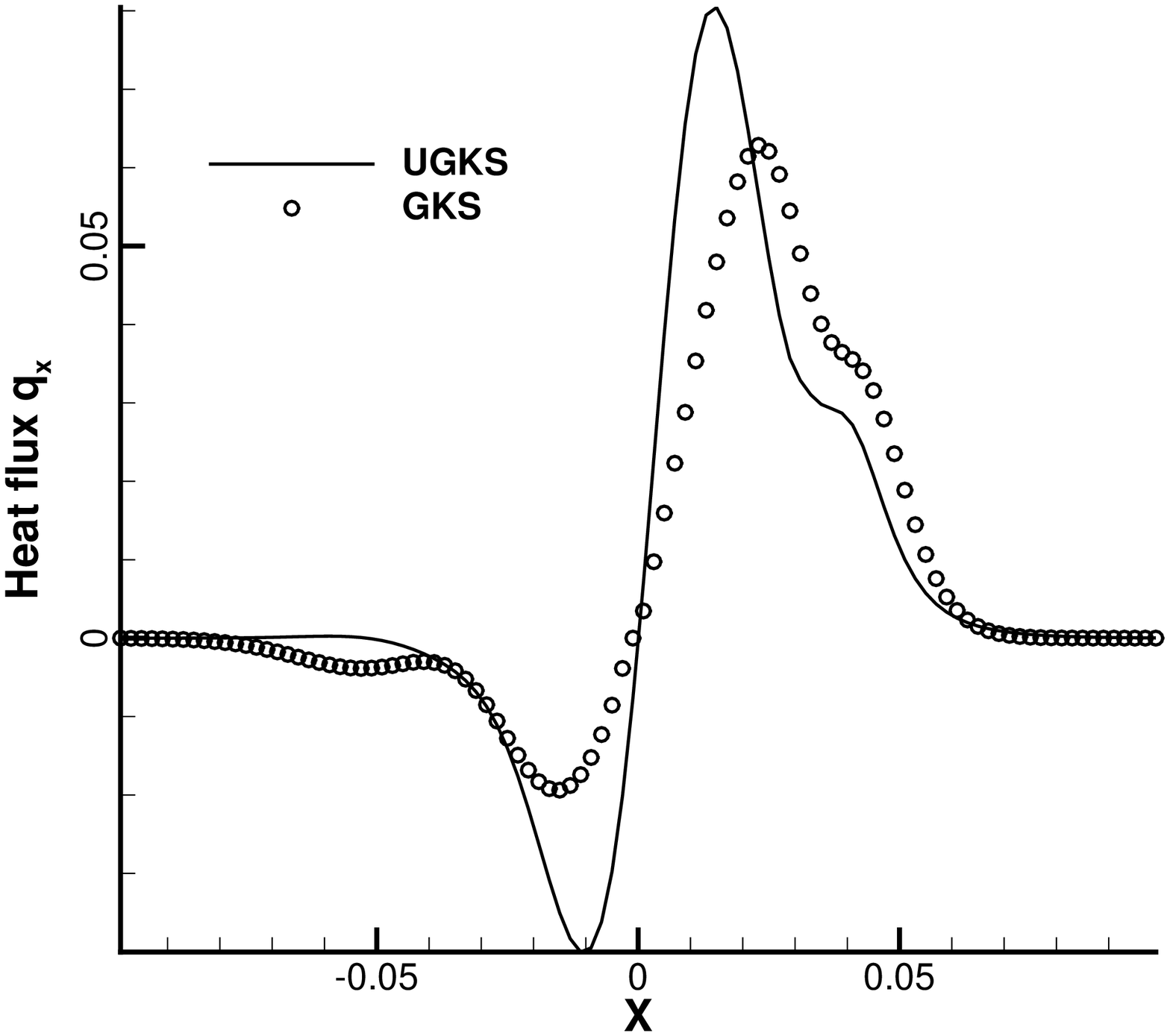}{e}
\includegraphics[width=0.4\textwidth]{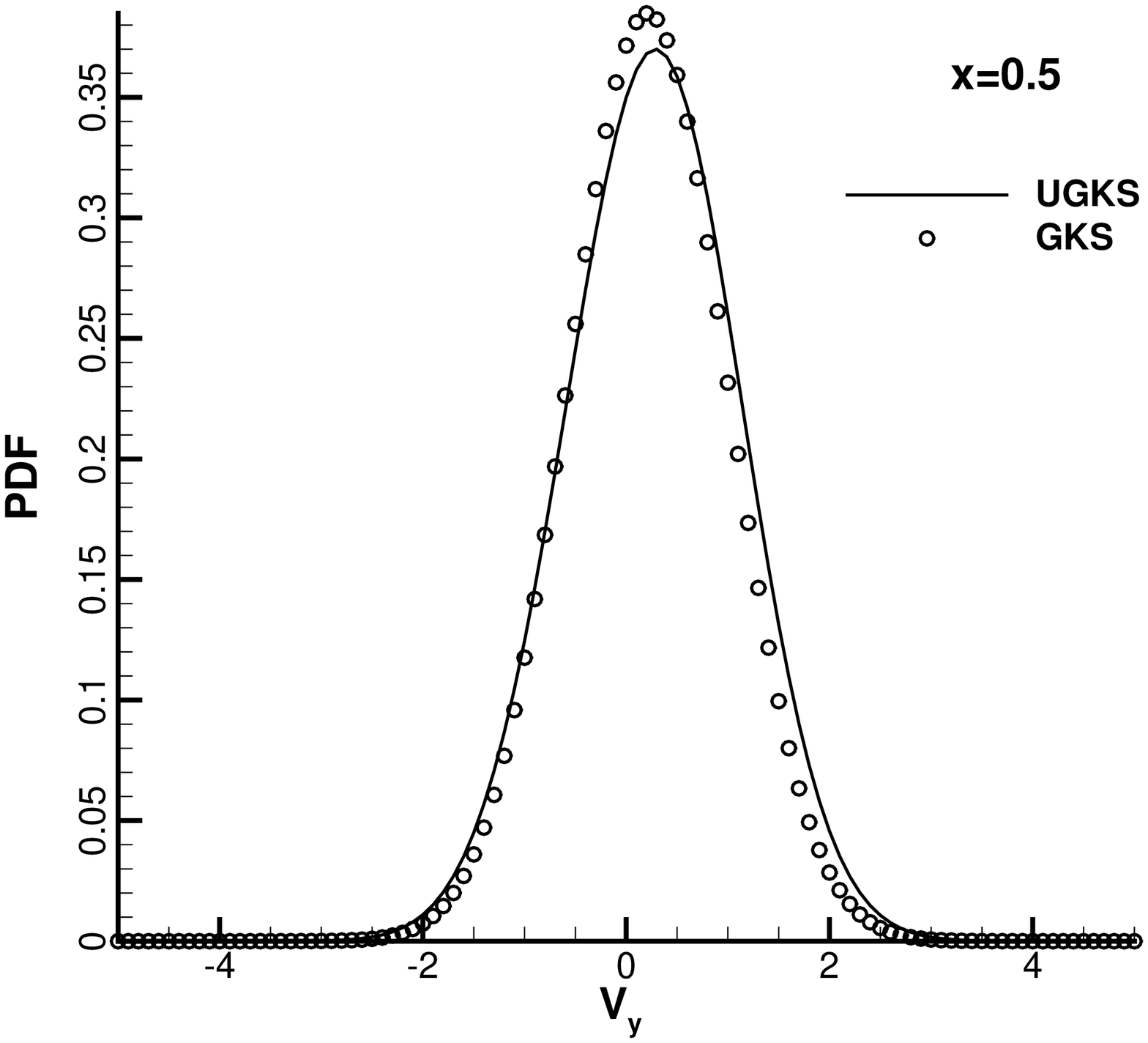}{f}
\caption{Results at $t=4\times 10^{-2}$ ($t/\tau=10.5$)
: a. density; b. x-velocity; c. y-velocity; d. temperature; e. x direction heat flux;
 f. velocity distribution at $x=0.5$.
For GKS $\Delta x/l_{mfp}=0.4$, $\Delta t/\tau=1.6\times 10^{-2}$,
and for UGKS $\Delta x/l_{mfp}=0.4$, $\Delta t/\tau=4\times 10^{-2}$.}\label{t2}
\end{figure}

\begin{figure}
\centering
\includegraphics[width=0.4\textwidth]{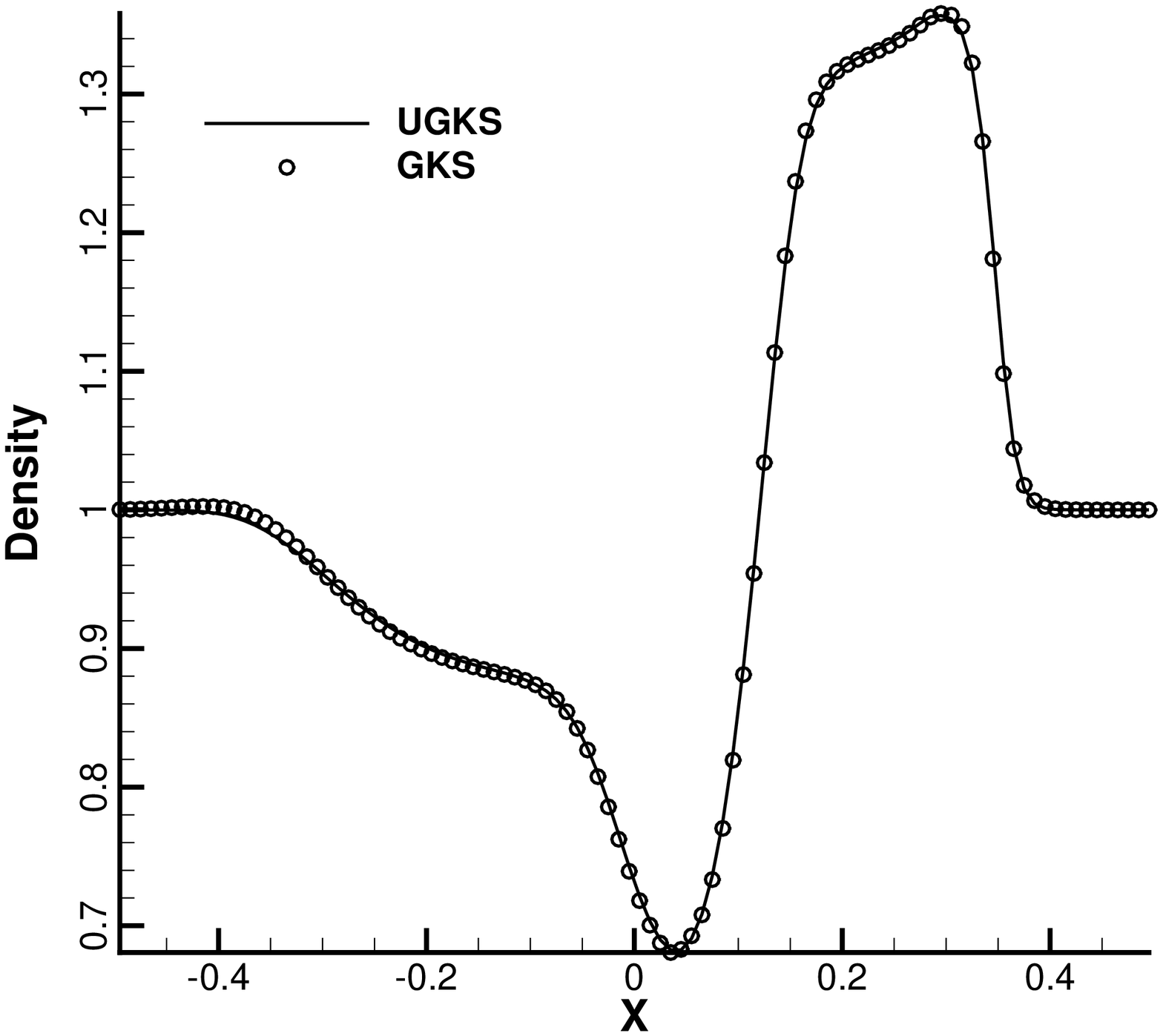}{a}
\includegraphics[width=0.4\textwidth]{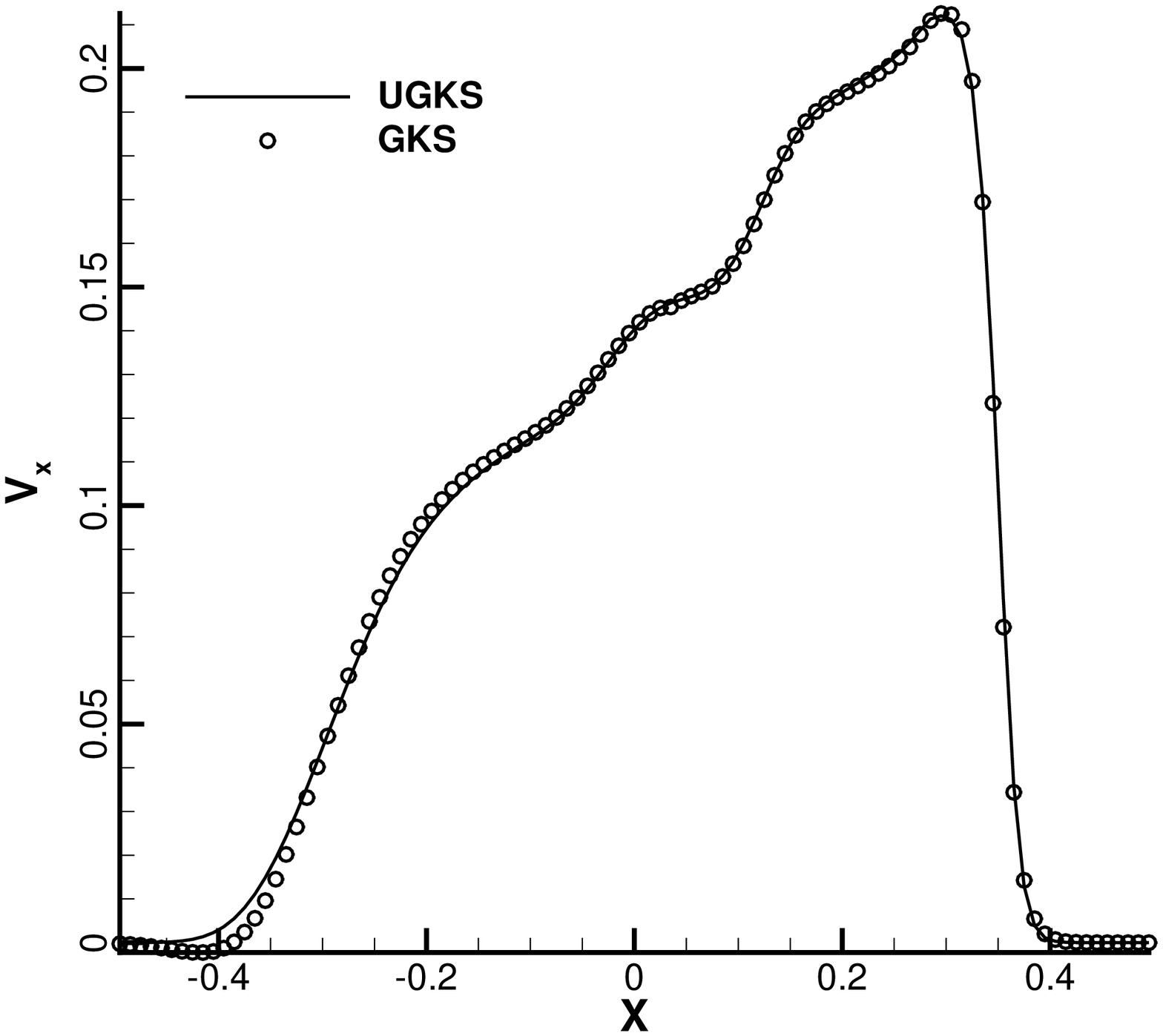}{b}\\
\includegraphics[width=0.4\textwidth]{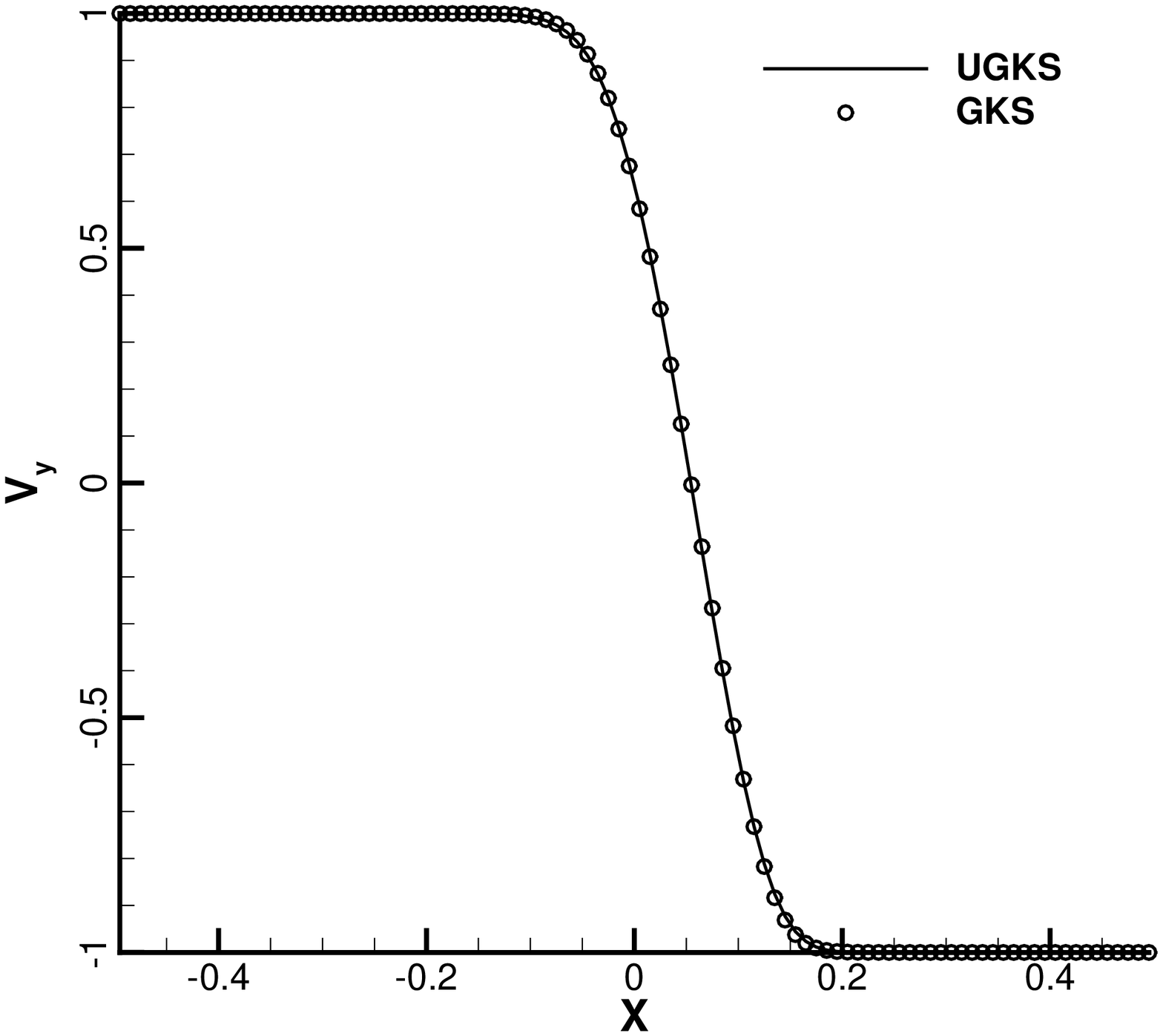}{c}
\includegraphics[width=0.4\textwidth]{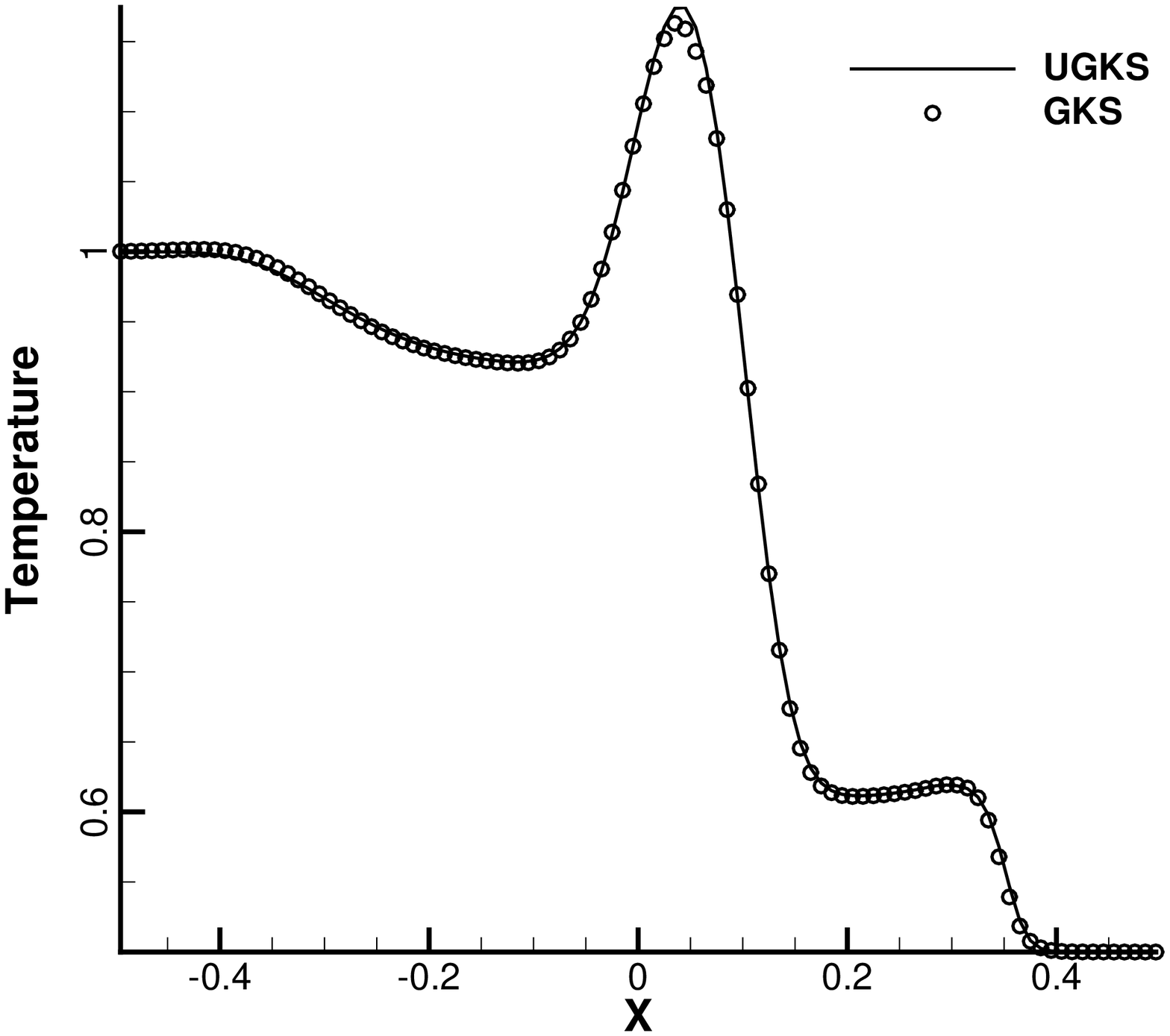}{d}\\
\includegraphics[width=0.4\textwidth]{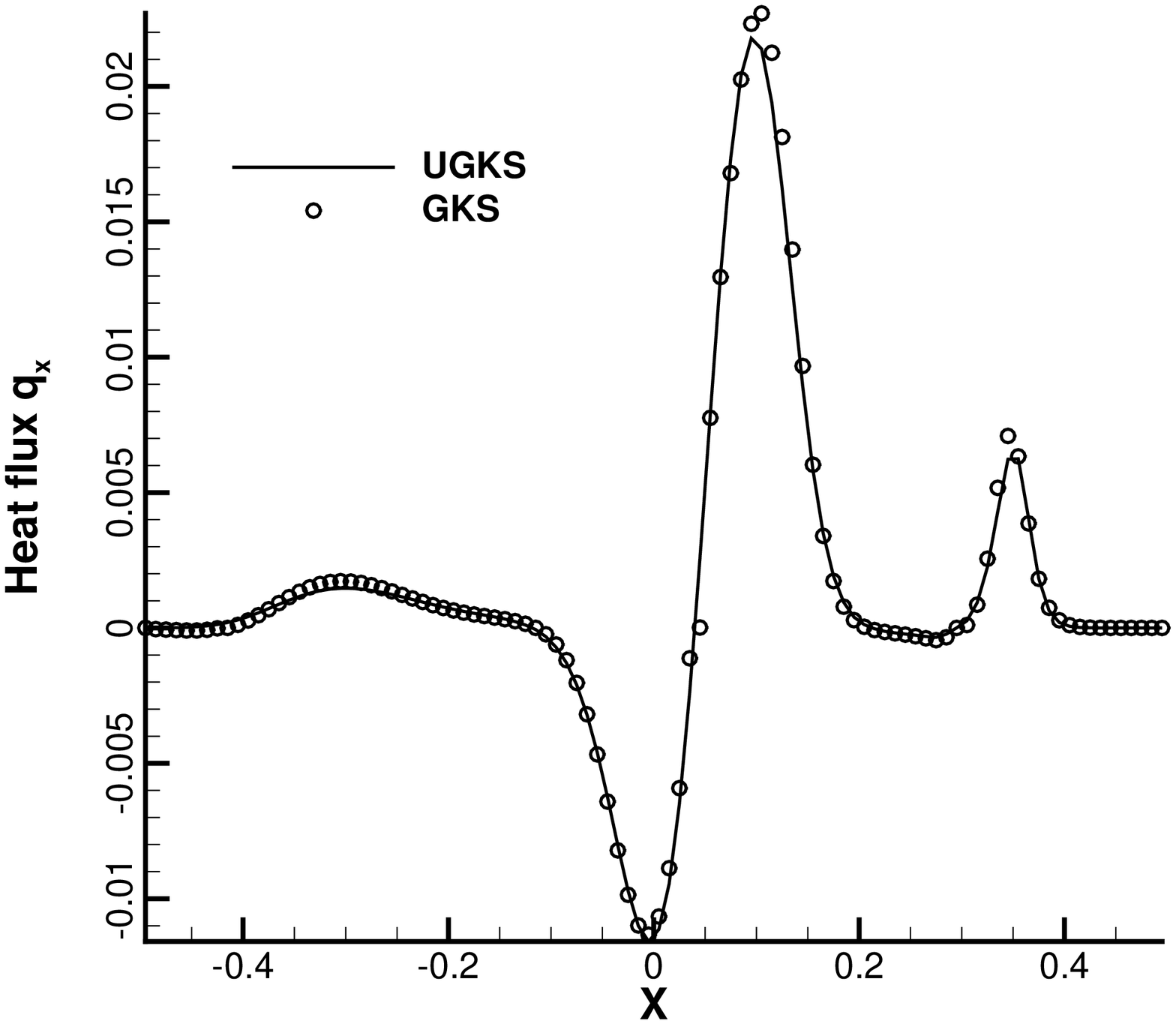}{e}
\includegraphics[width=0.4\textwidth]{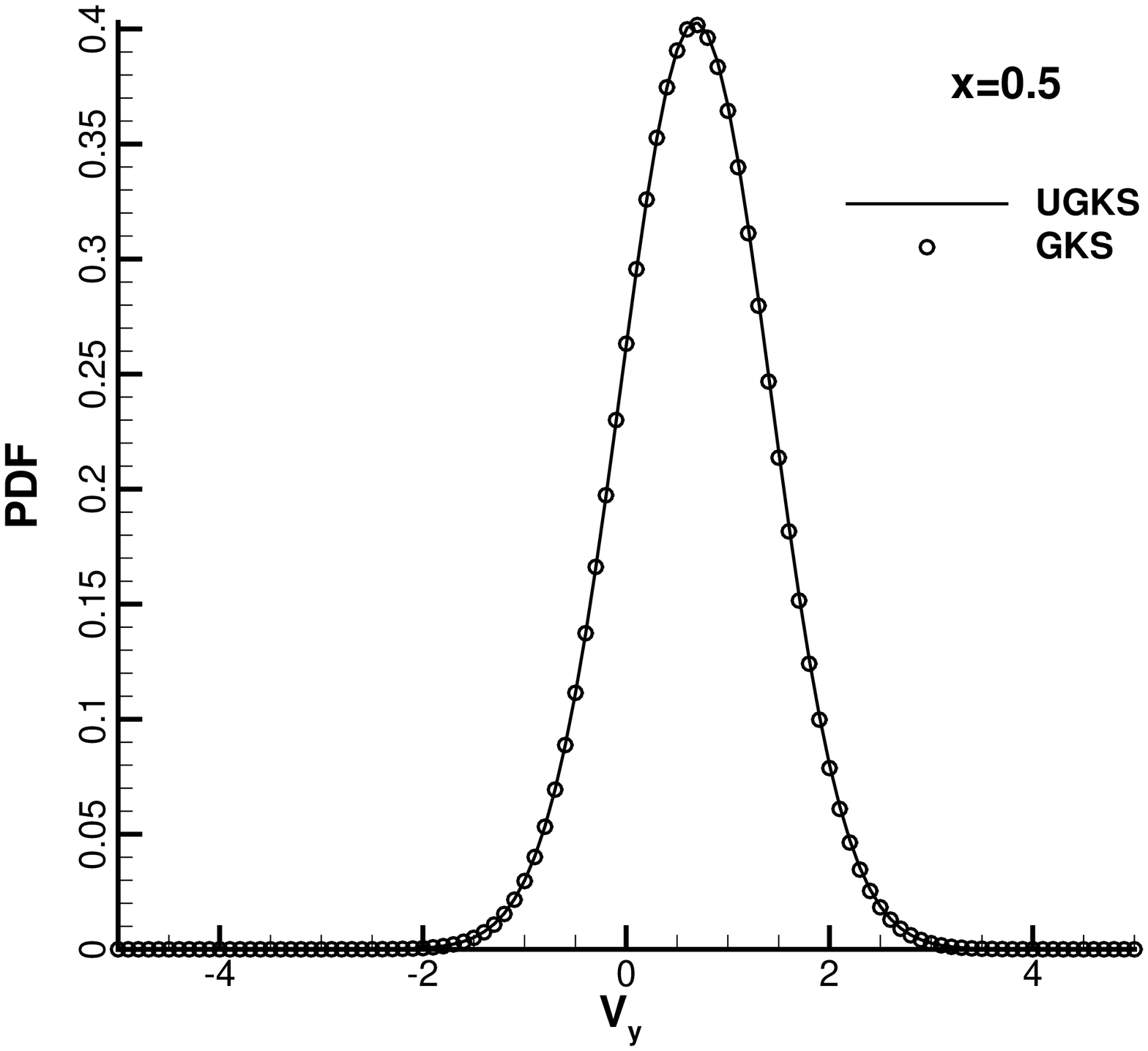}{f}
\caption{Results at $t=4\times 10^{-1}$ ($t/\tau=104.5$)
: a. density; b. x-velocity; c. y-velocity; d. temperature; e. x direction heat flux;
 f. velocity distribution at $x=0.5$.
 For GKS $\Delta x/l_{mfp}=2$, $\Delta t/\tau=0.4$,
 and for UGKS $\Delta x/l_{mfp}=2$, $\Delta t/\tau=0.2$.}\label{t1}
\end{figure}

\begin{figure}
\centering
\includegraphics[width=0.4\textwidth]{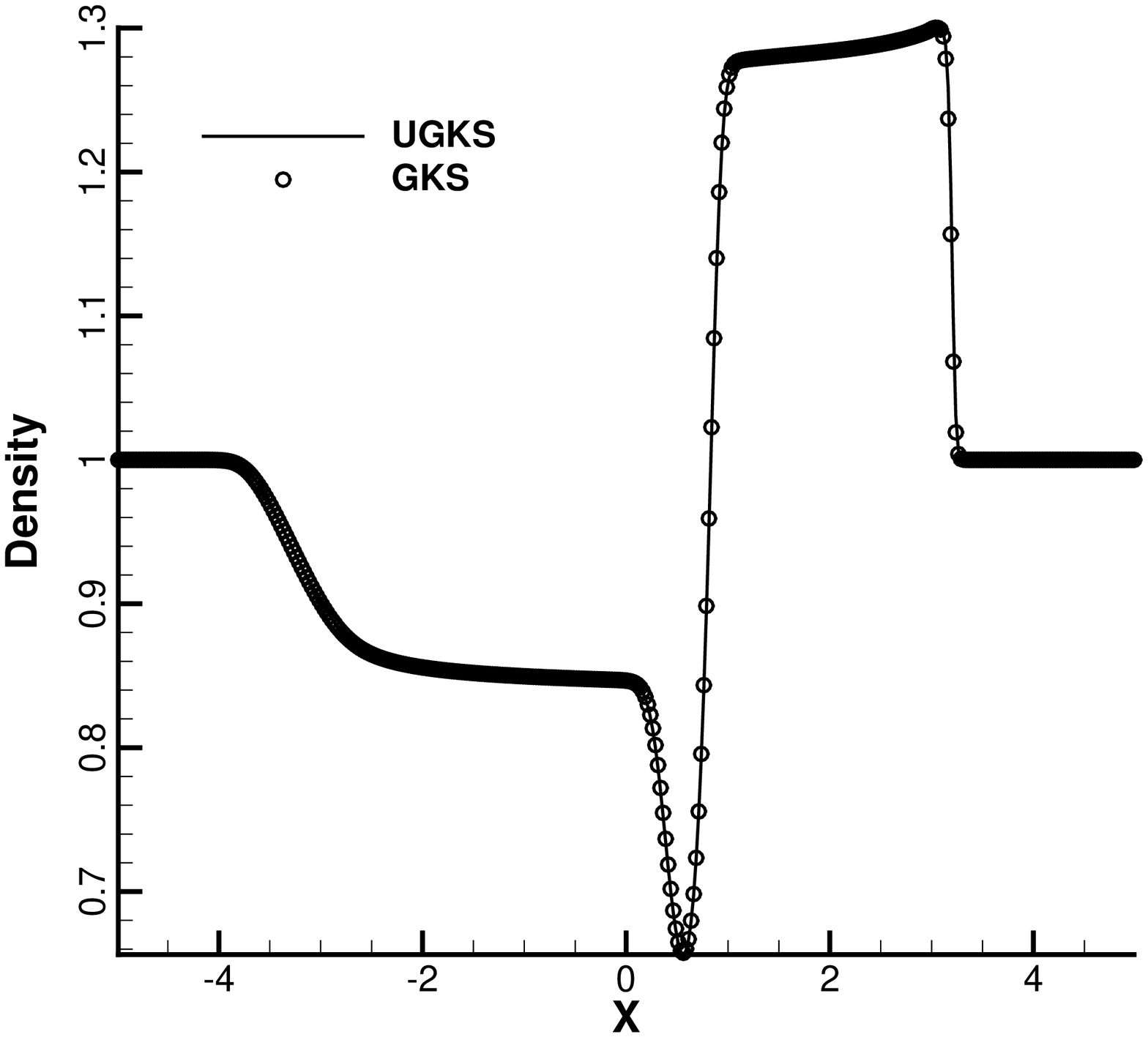}{a}
\includegraphics[width=0.4\textwidth]{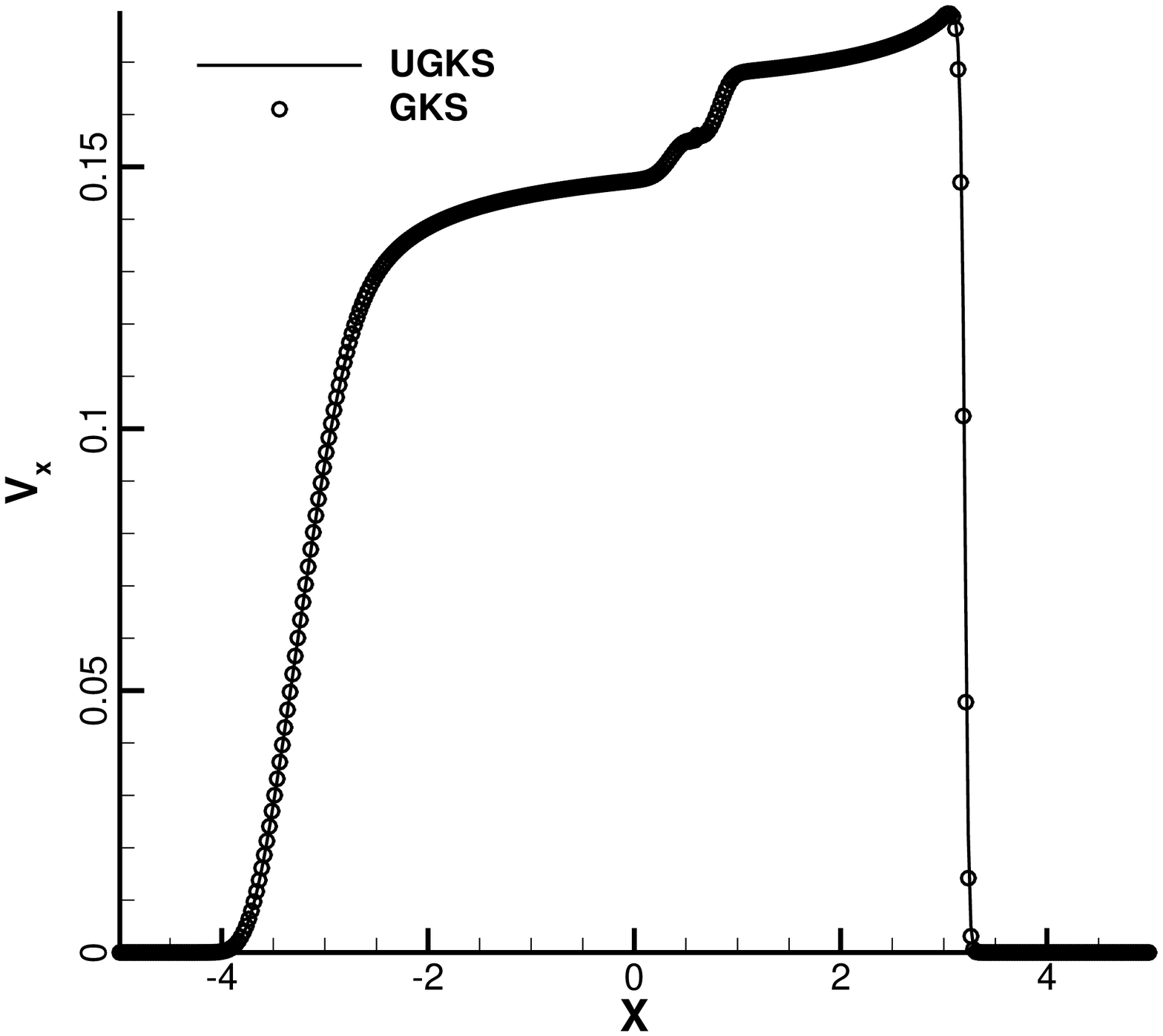}{b}\\
\includegraphics[width=0.4\textwidth]{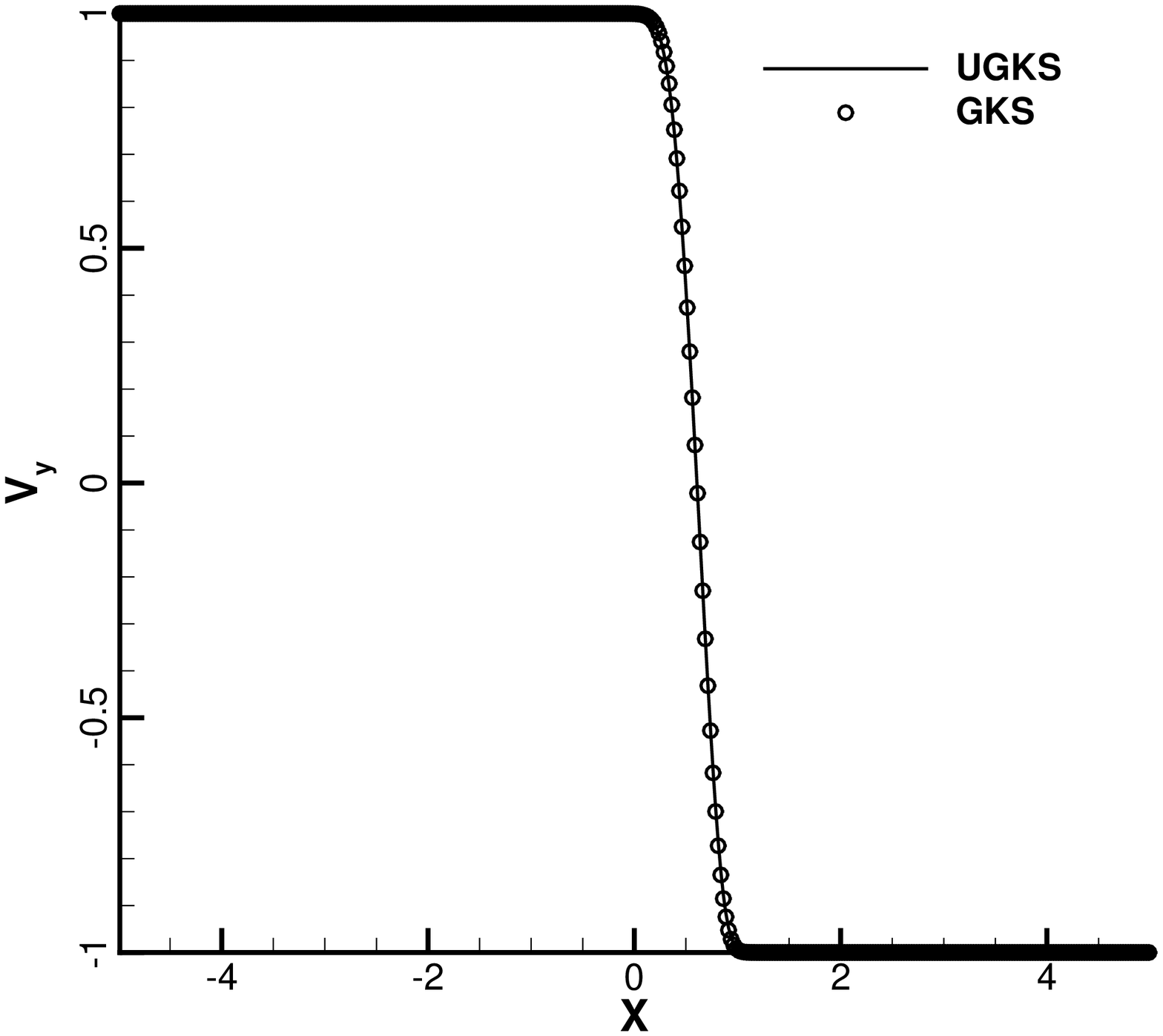}{c}
\includegraphics[width=0.4\textwidth]{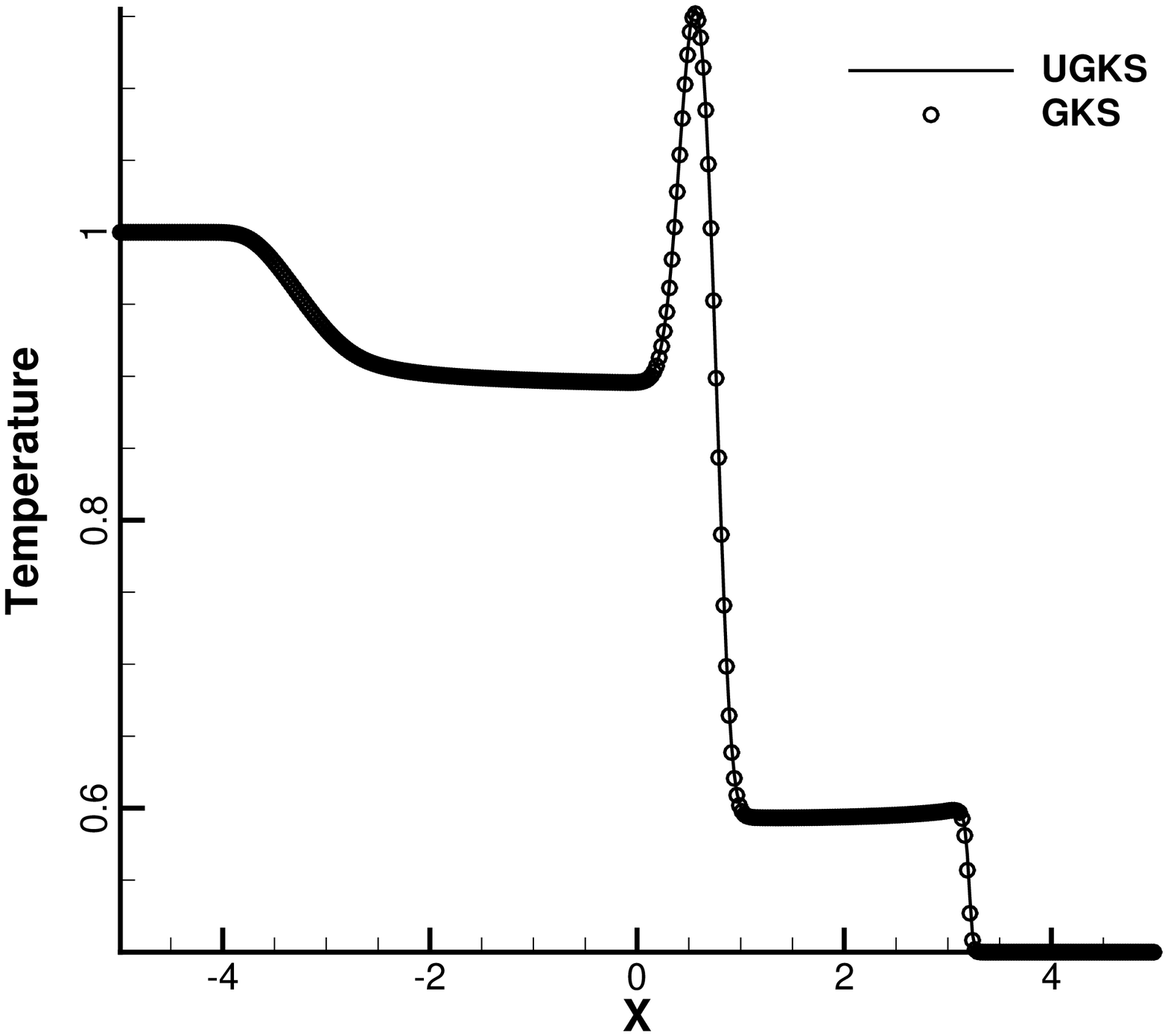}{d}\\
\includegraphics[width=0.4\textwidth]{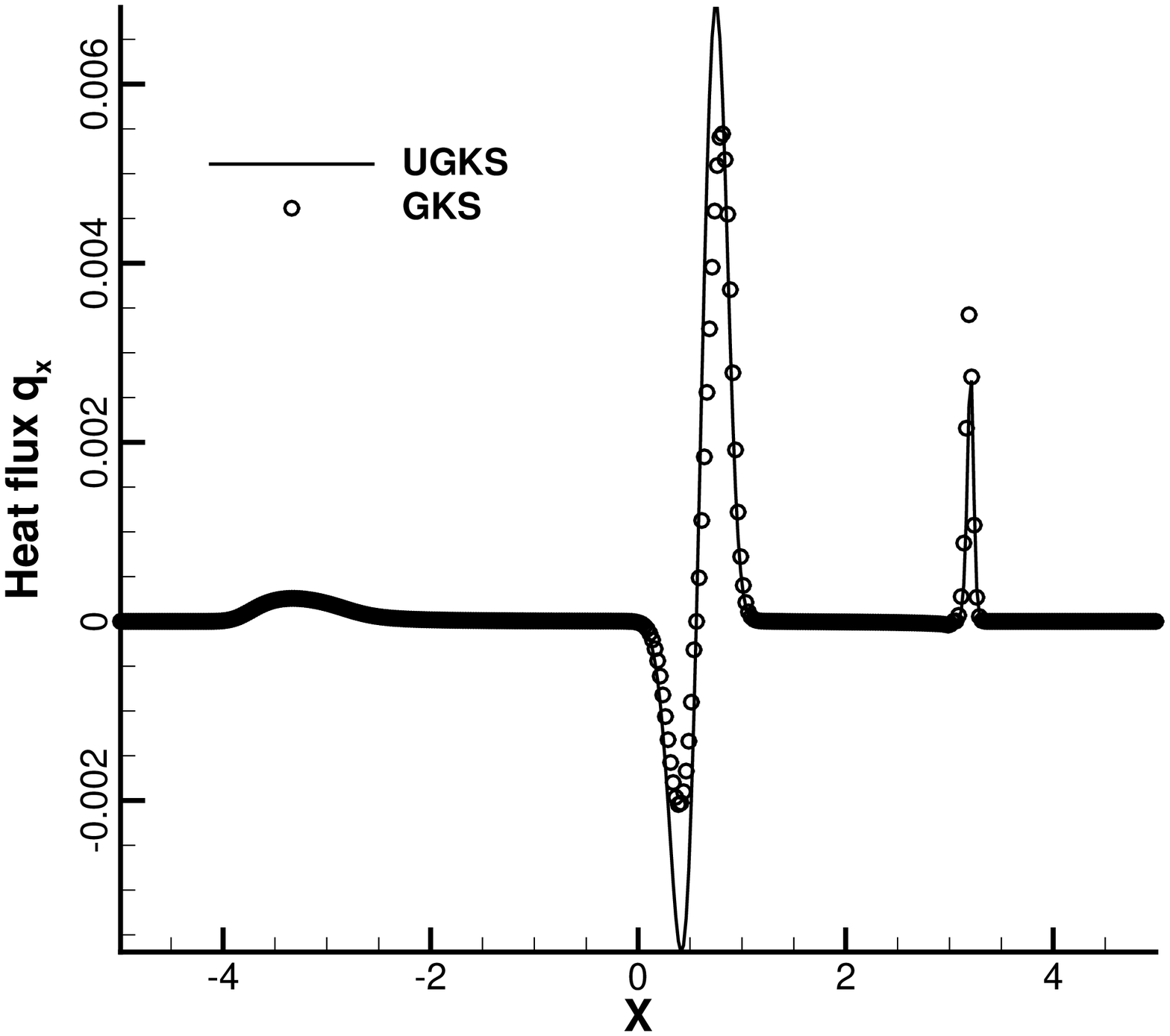}{e}
\includegraphics[width=0.4\textwidth]{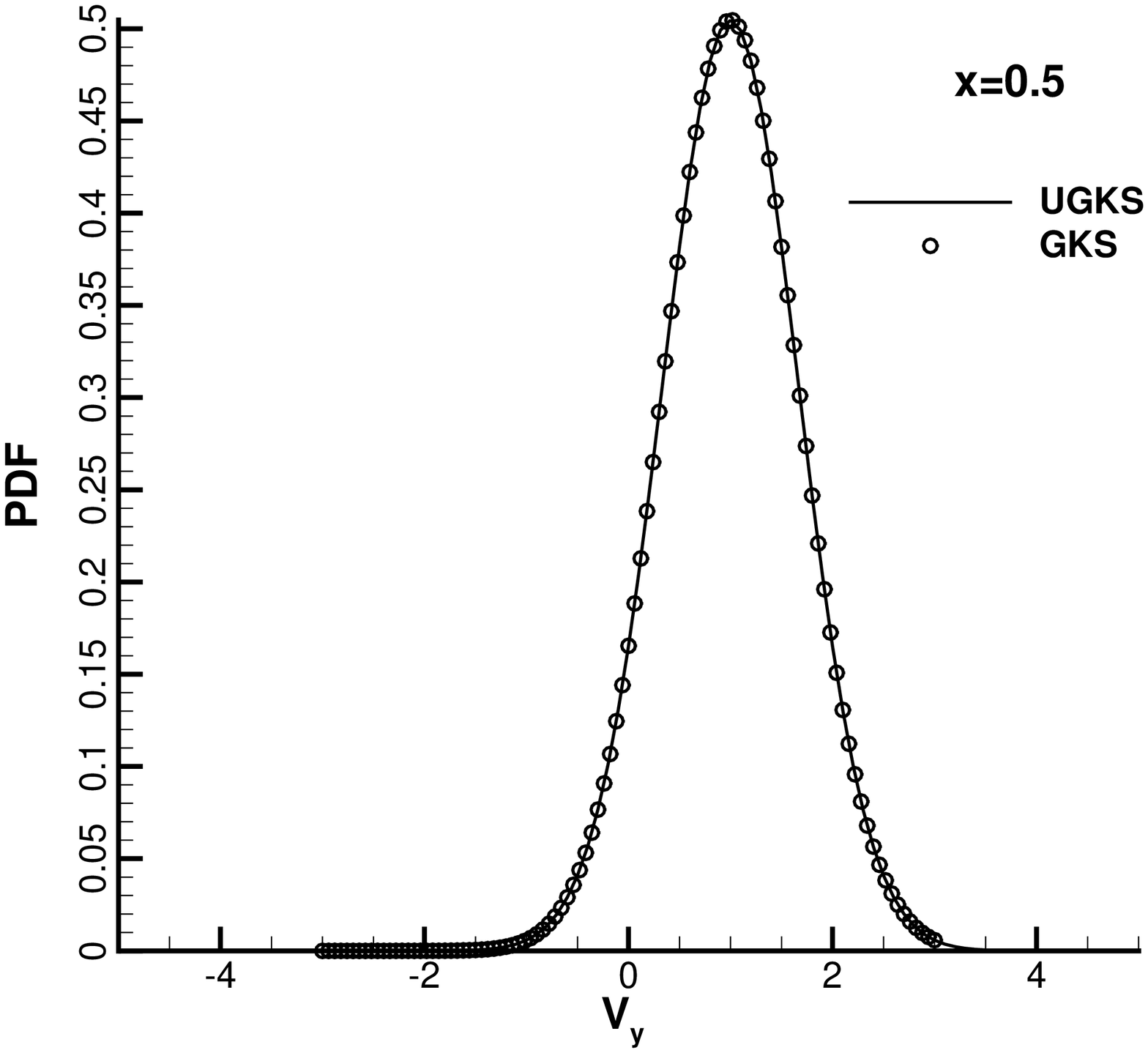}{f}
\caption{Results at $t=4$ ($t/\tau=1.05\times10^3$)
: a. density; b. x-velocity; c. y-velocity; d. temperature; e. x direction heat flux;
f. velocity distribution at $x=0.5$.
For GKS $\Delta x/l_{mfp}=10$, $\Delta t/\tau=2$,
and for UGKS $\Delta x/l_{mfp}=10$, $\Delta t/\tau=1$.}\label{t0}
\end{figure}

\begin{figure}
\centering
\includegraphics[width=0.4\textwidth]{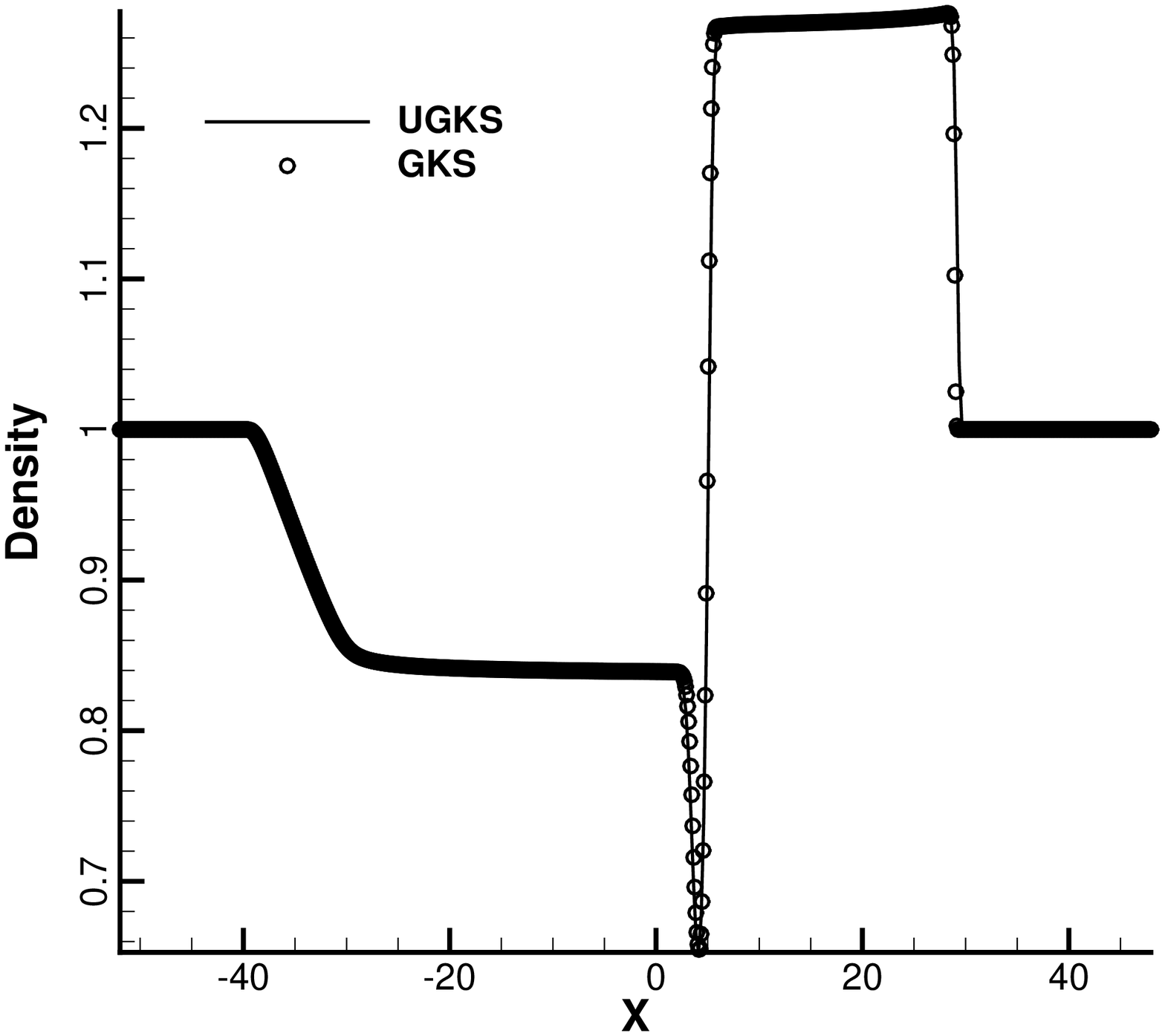}{a}
\includegraphics[width=0.4\textwidth]{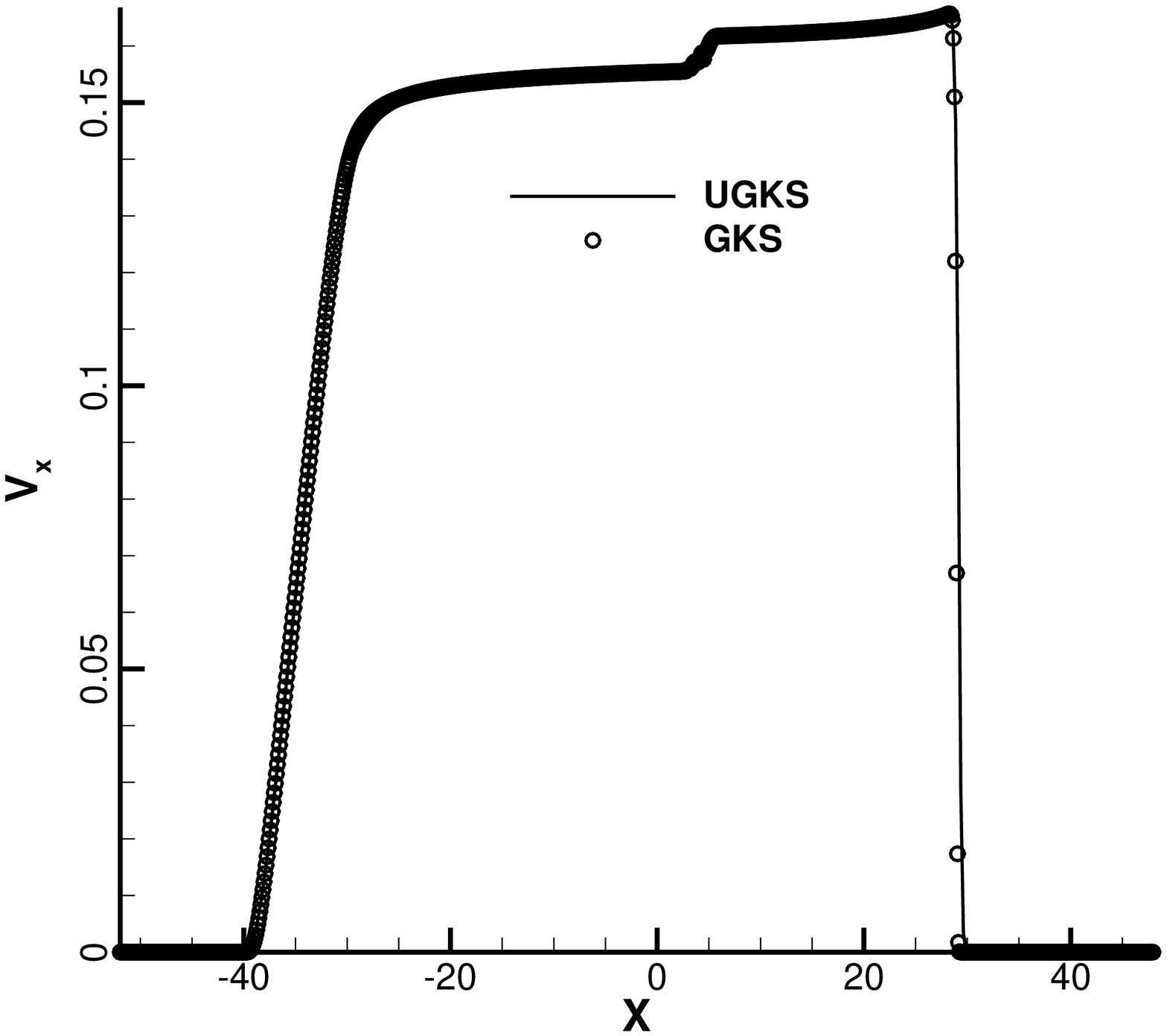}{b}\\
\includegraphics[width=0.4\textwidth]{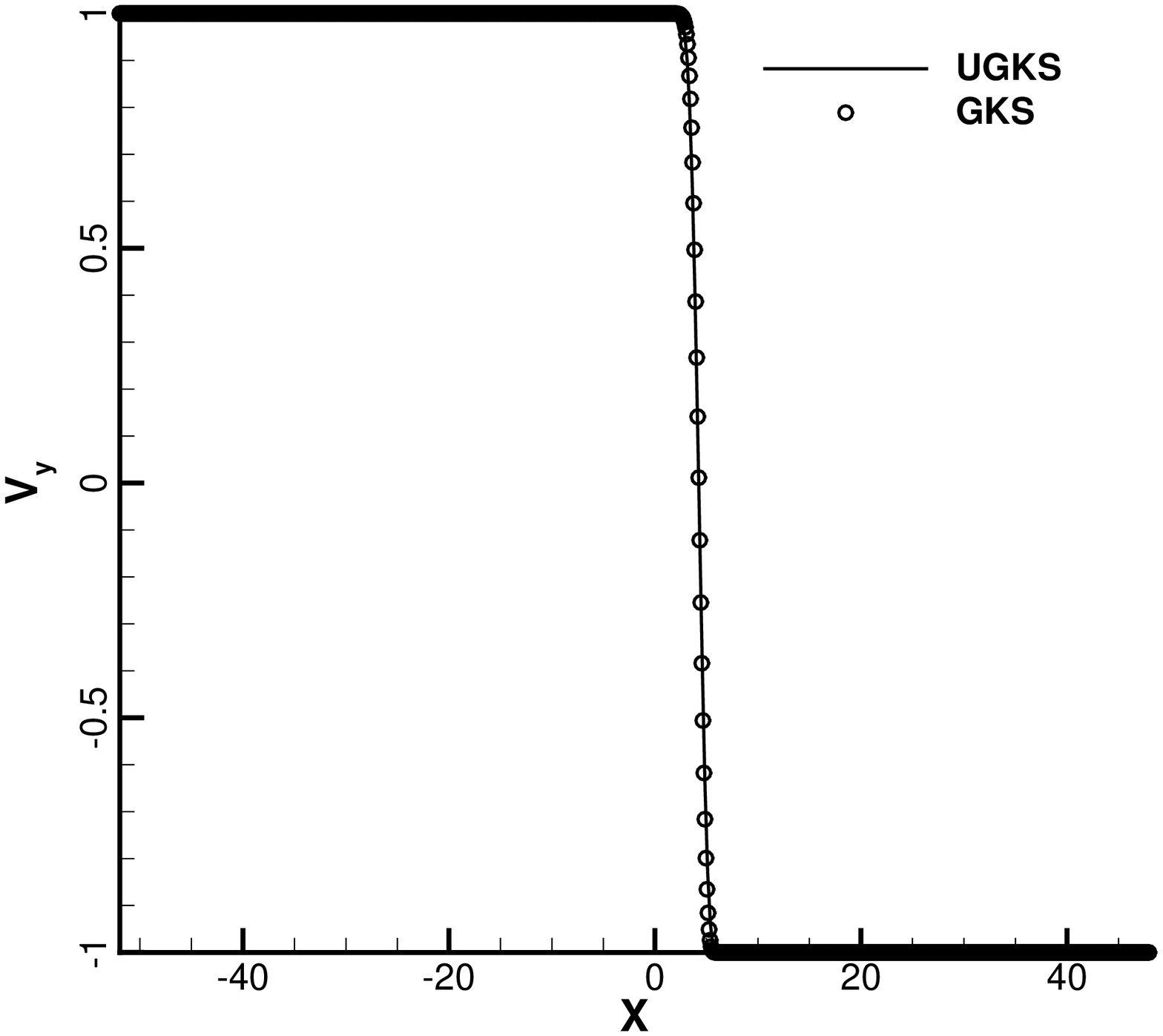}{c}
\includegraphics[width=0.4\textwidth]{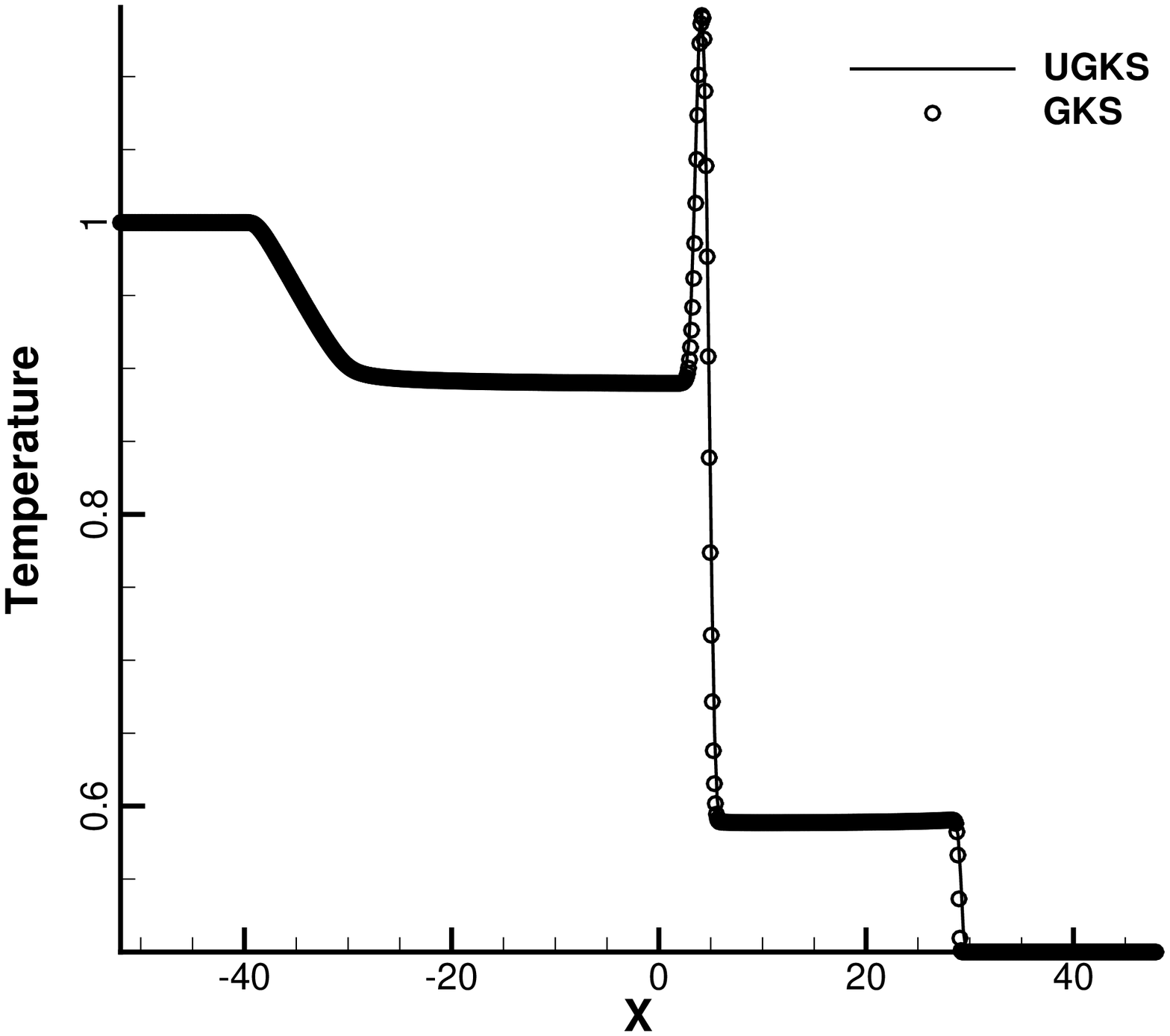}{d}\\
\includegraphics[width=0.4\textwidth]{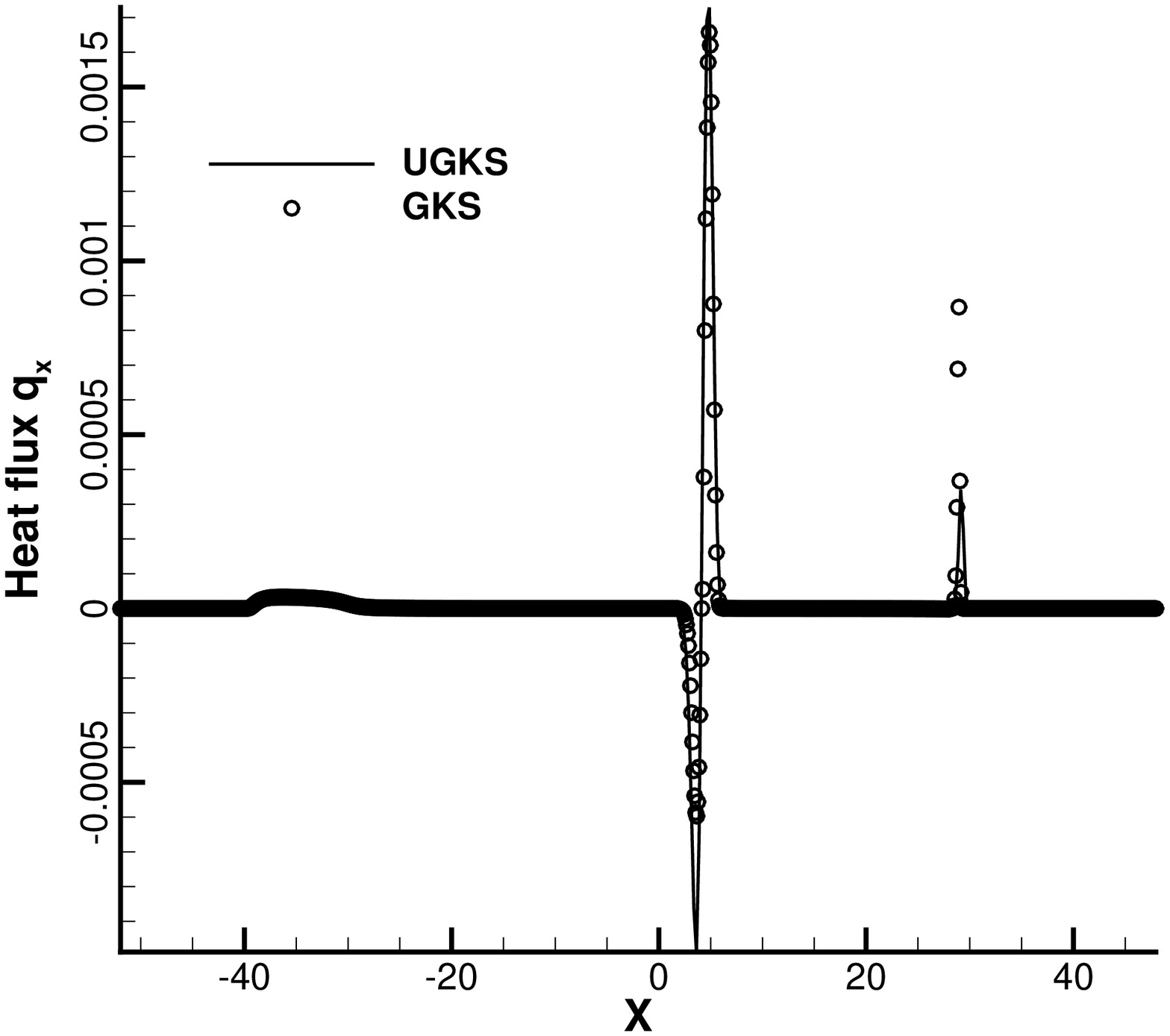}{e}
\includegraphics[width=0.4\textwidth]{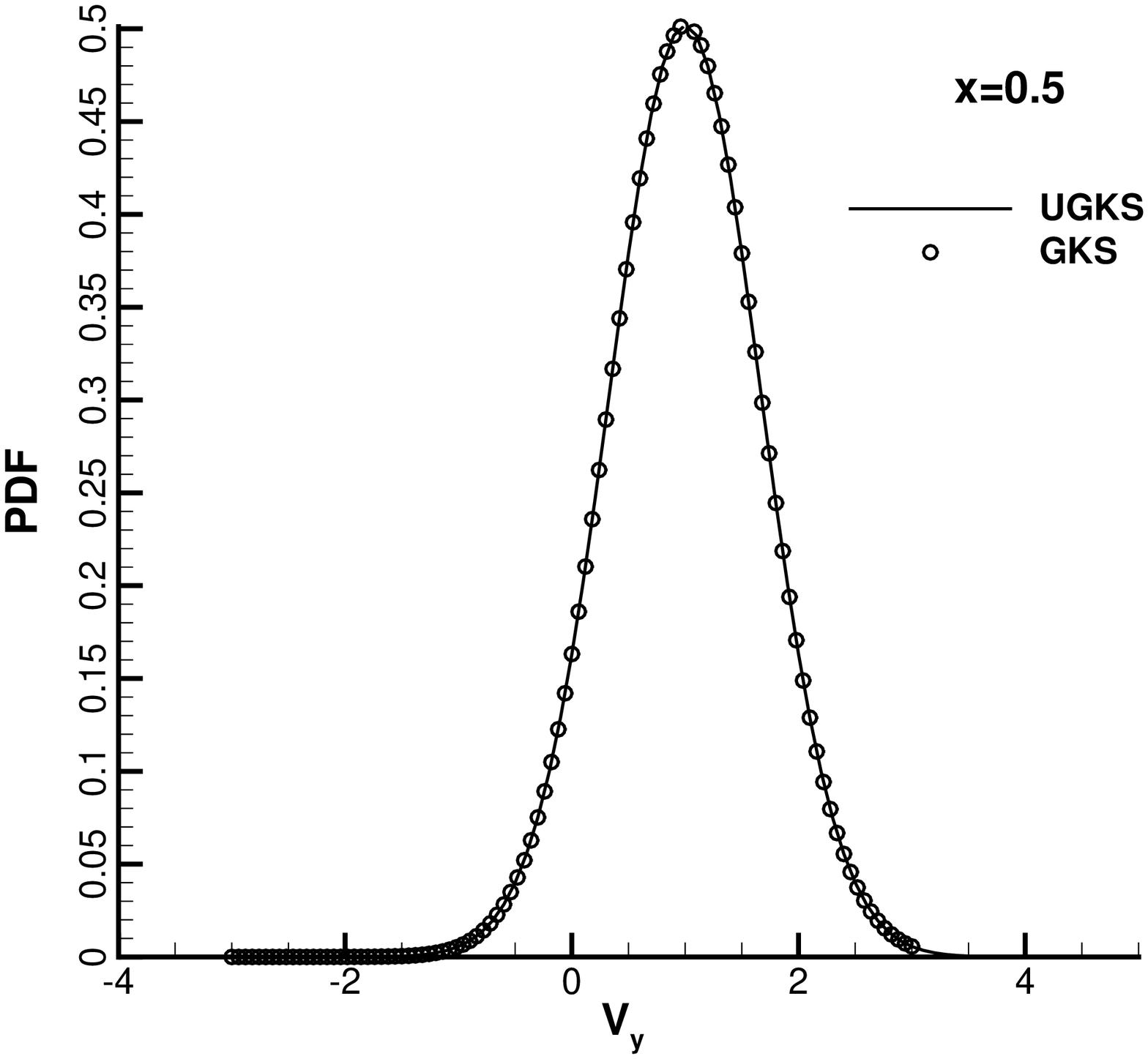}{f}
\caption{Results at $t=40$ ($t/\tau=1.05\times10^4$)
: a. density; b. x-velocity; c. y-velocity; d. temperature; e. x direction heat flux;
f. velocity distribution at $x=0.5$.
For GKS $\Delta x/l_{mfp}=40$, $\Delta t/\tau=7$,
and for UGKS $\Delta x/l_{mfp}=100$, $\Delta t/\tau=10$.}\label{t-1}
\end{figure}

\begin{figure}
\centering
\includegraphics[width=0.4\textwidth]{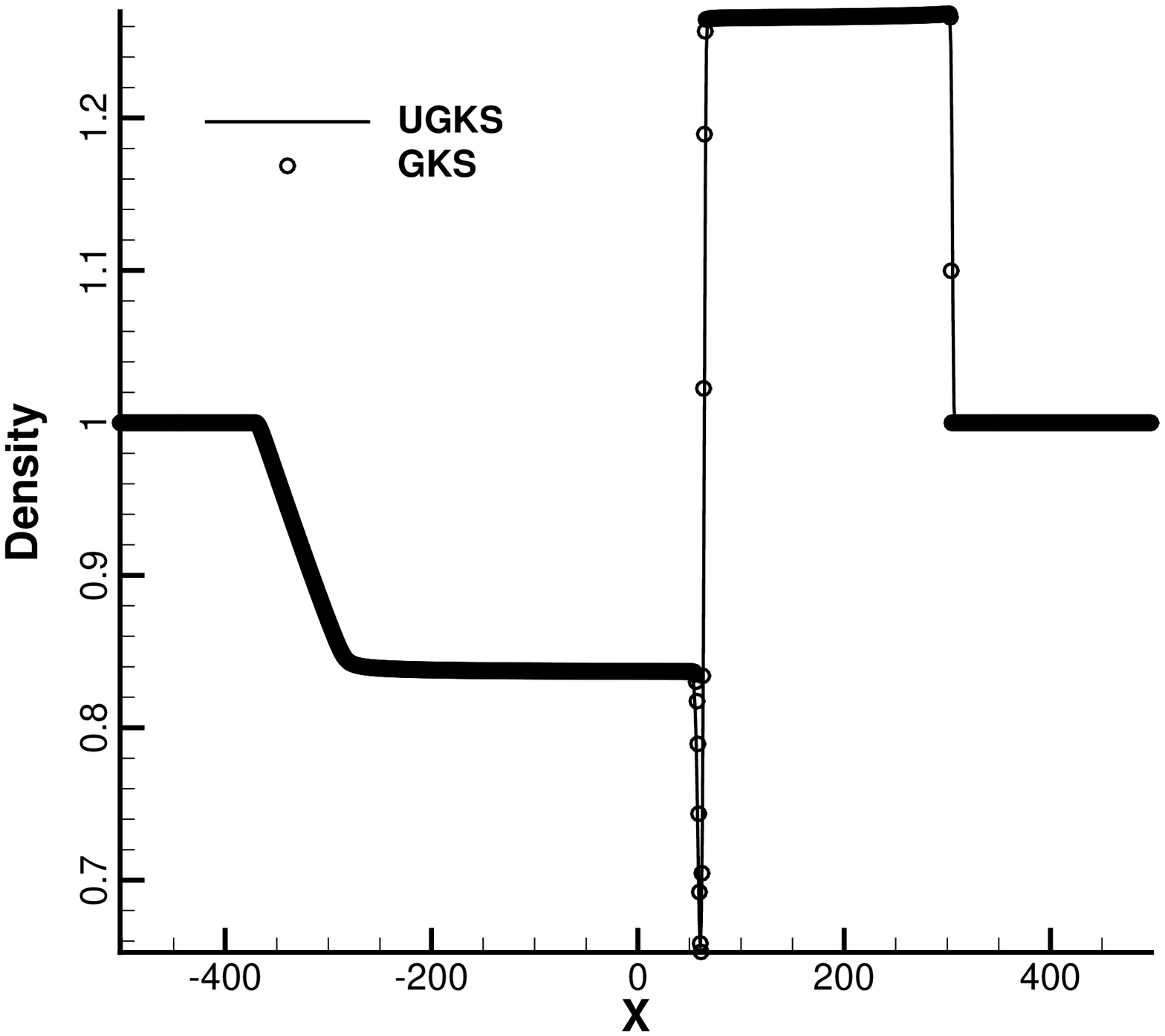}{a}
\includegraphics[width=0.4\textwidth]{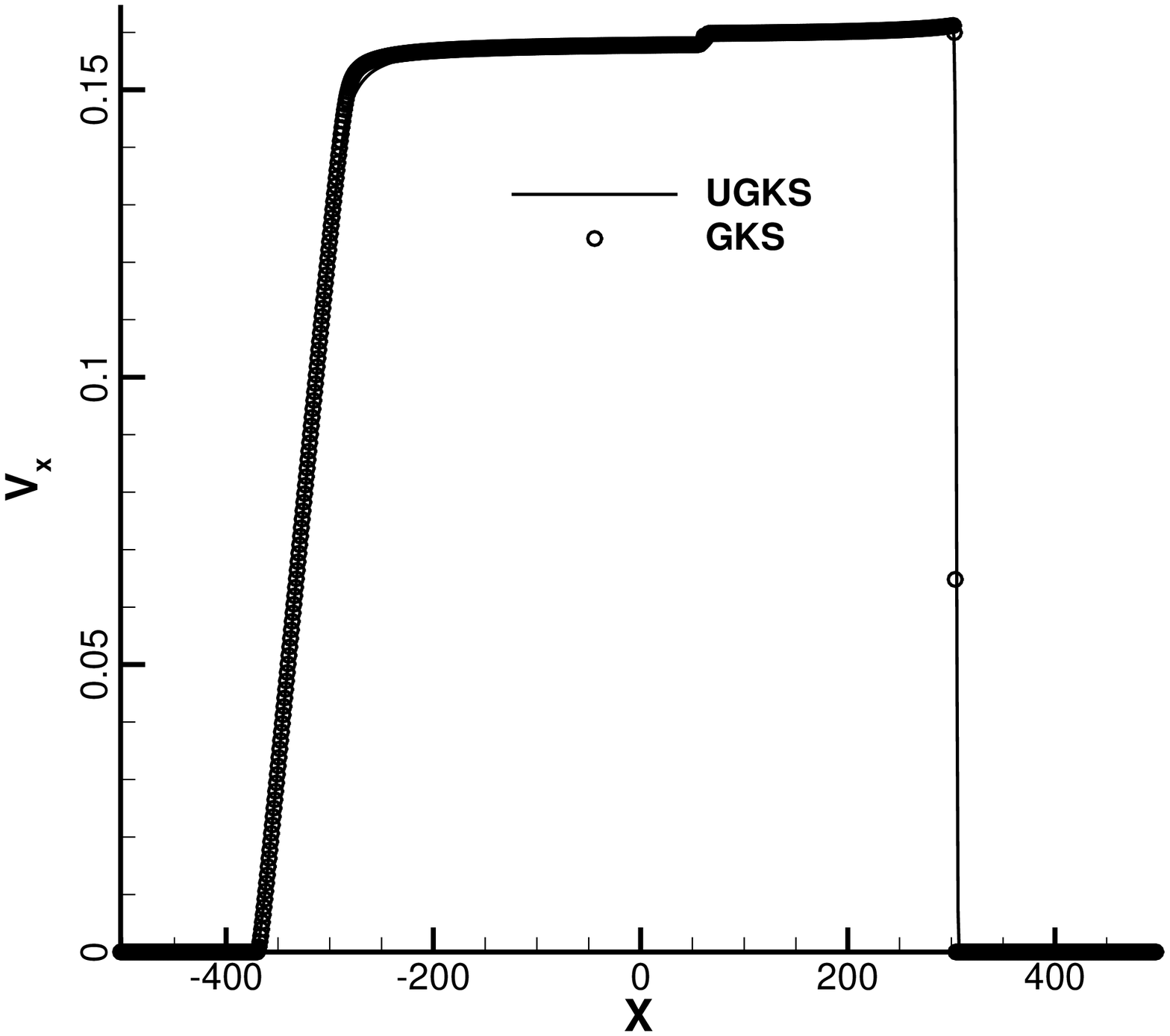}{b}\\
\includegraphics[width=0.4\textwidth]{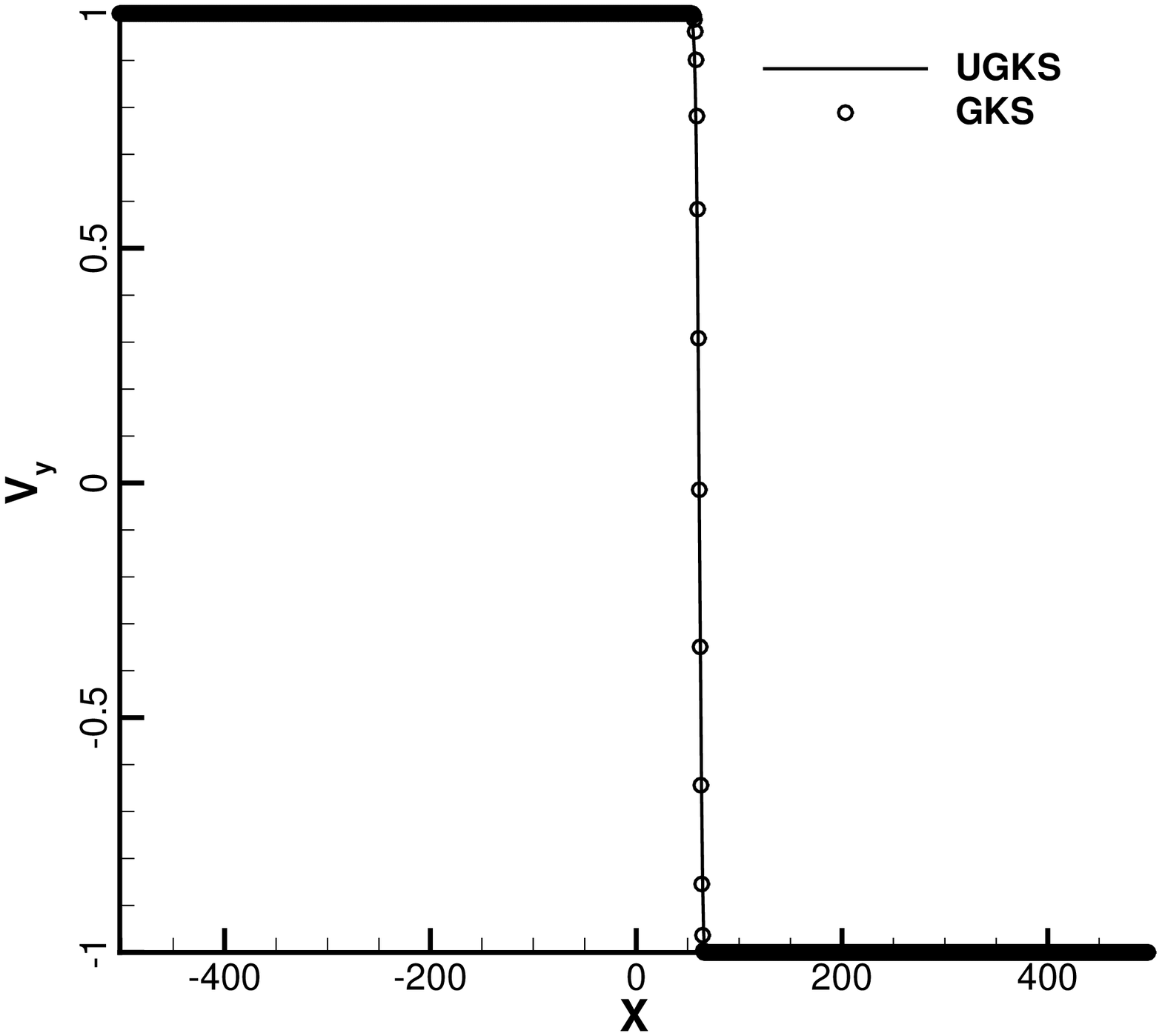}{c}
\includegraphics[width=0.4\textwidth]{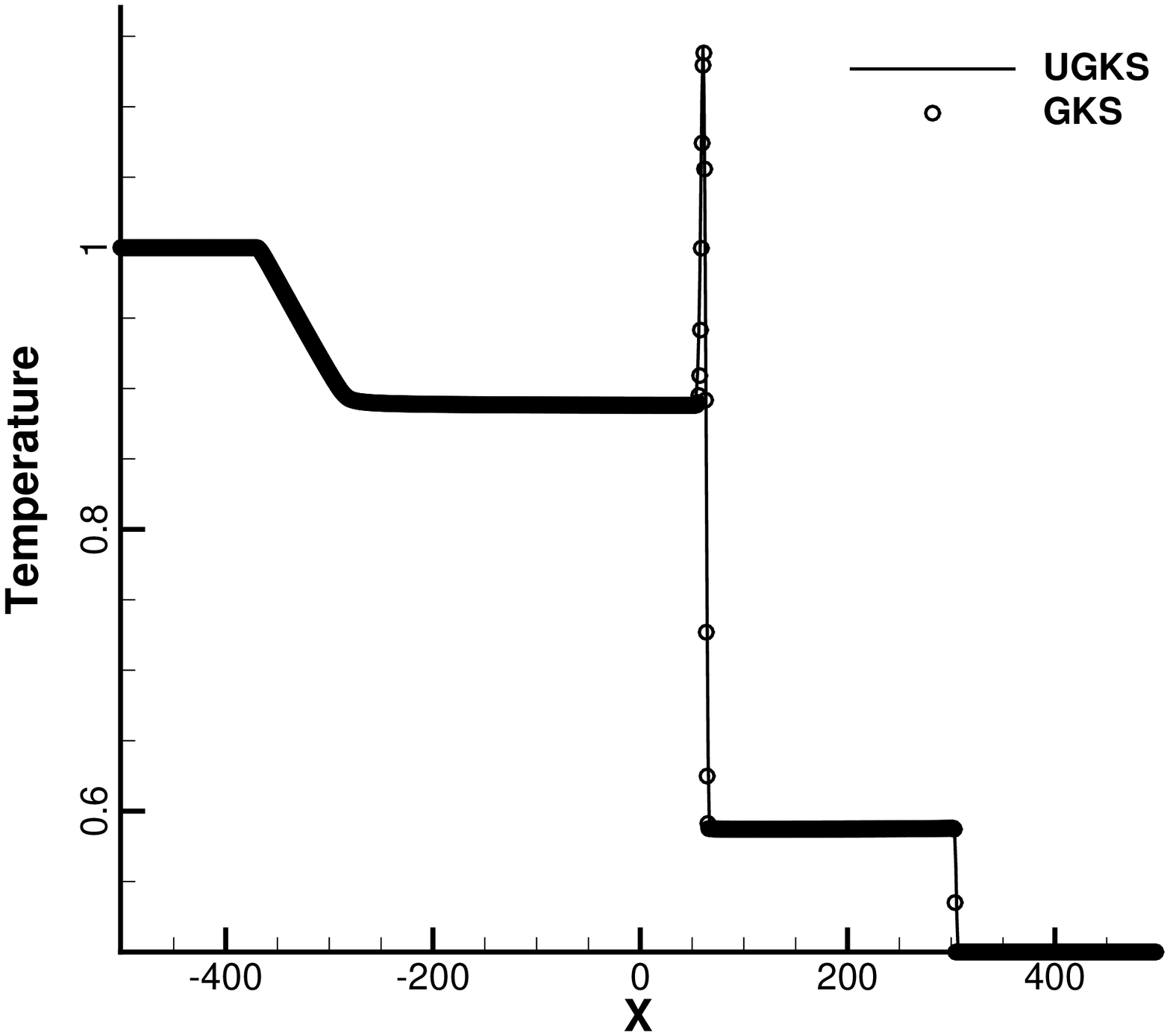}{d}\\
\includegraphics[width=0.4\textwidth]{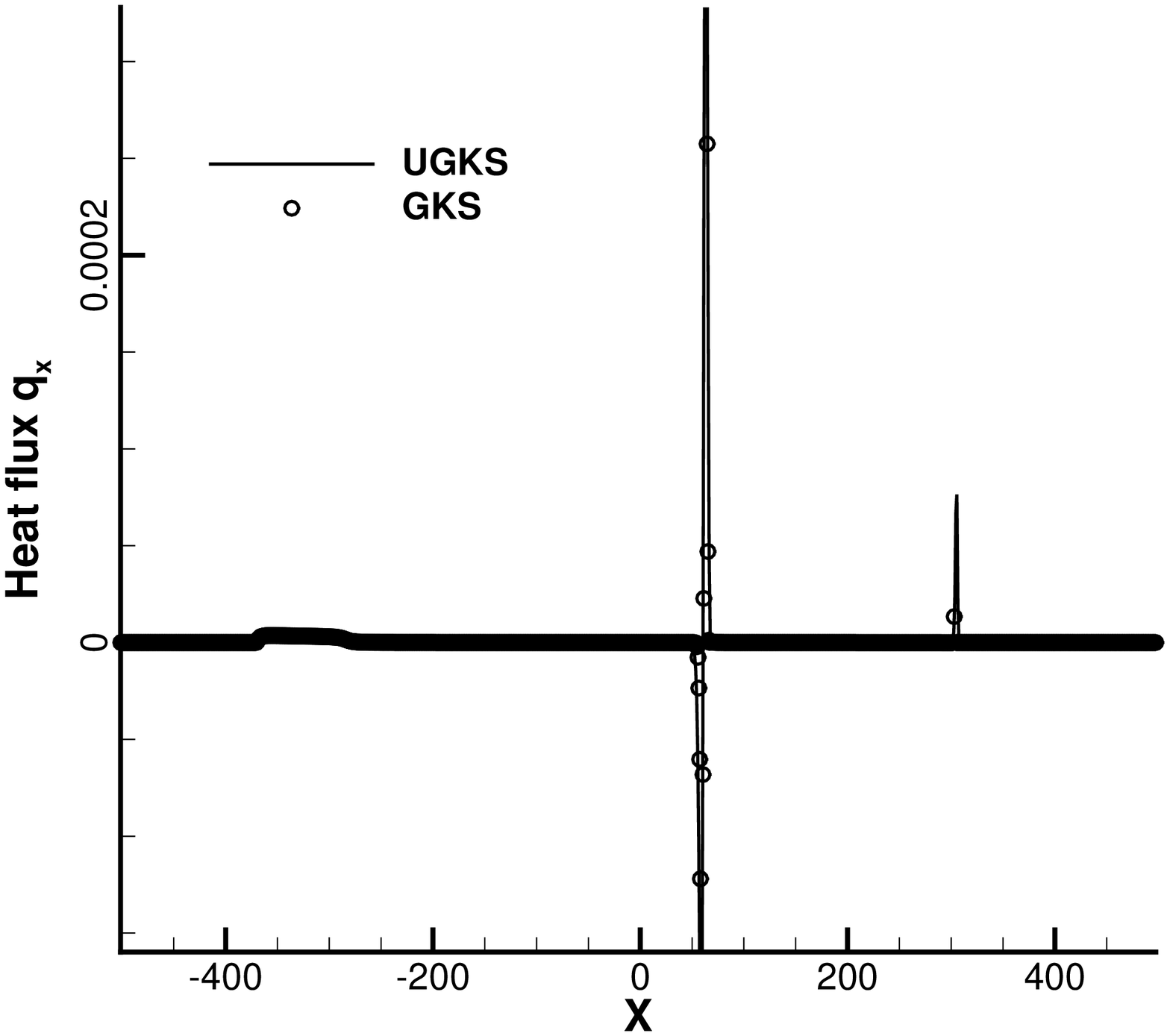}{e}
\includegraphics[width=0.4\textwidth]{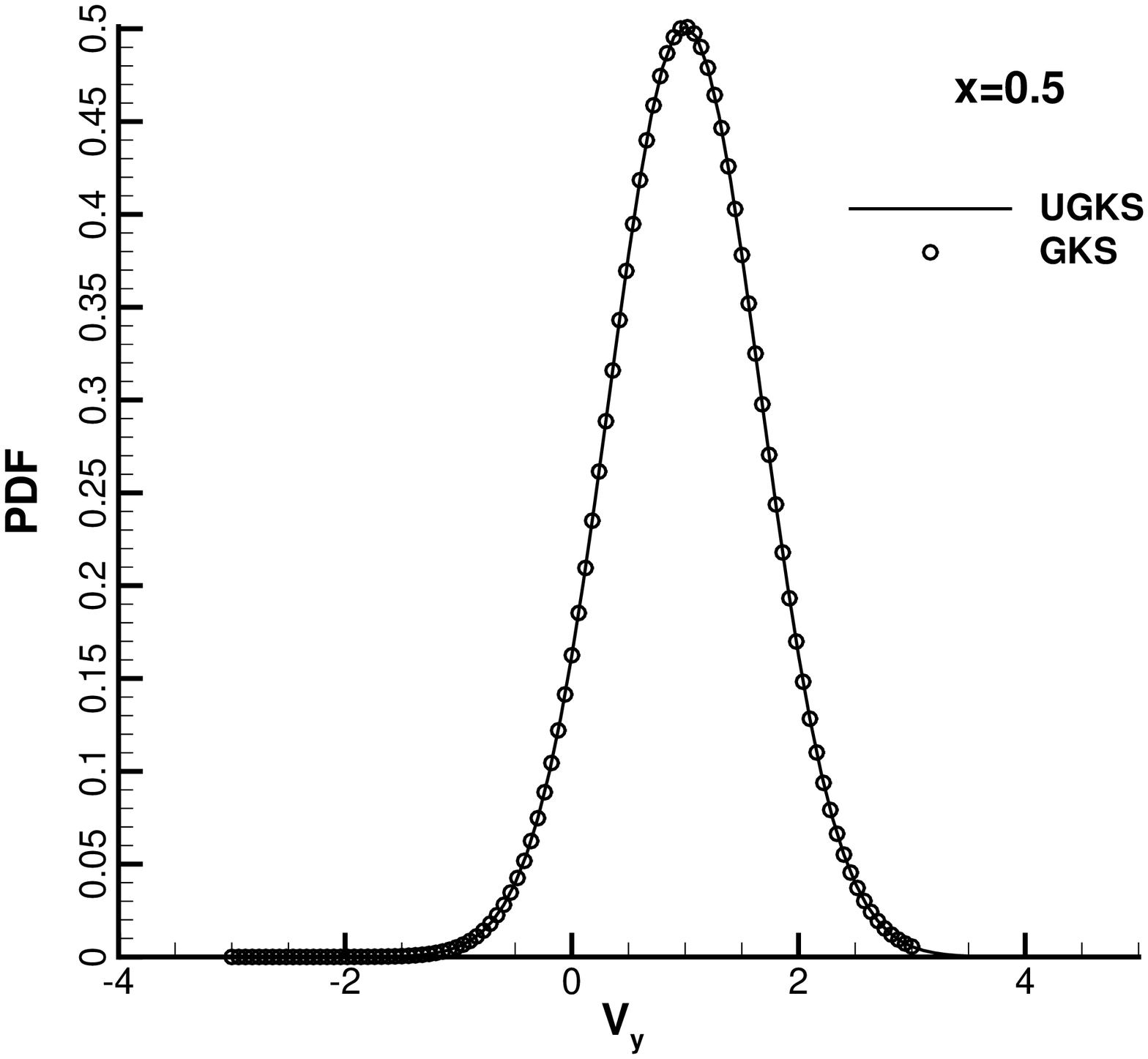}{f}
\caption{Results at $t=400$ ($t/\tau=1.05\times10^5$)
: a. density; b. x-velocity; c. y-velocity; d. temperature; e. x direction heat flux;
f. velocity distribution at $x=0.5$.
For GKS $\Delta x/l_{mfp}=80$, $\Delta t/\tau=10$ (1250 symbols plotted),
and for UGKS $\Delta x/l_{mfp}=500$, $\Delta t/\tau=50$.}\label{t-2}
\end{figure}

We plot the density, velocity, temperature, heat flux,
as well as the velocity distribution functions at time
$t_1=4\times 10^{-3}$, $t_2=4\times 10^{-2}$, $t_3=0.4$, $t_4=4$, $t_5=40$, and $t_6=400$ with a changeable cell size in order to identify the shear solution in different scales.
For GKS, the cell number in x direction is $100$ for Fig. \ref{t3}-\ref{t1}, $400$ for Fig.\ref{t0}, $1000$ for Fig.\ref{t-1}, and $5000$ for Fig. \ref{t-2}.
For UGKS, the cell number in x direction is $100$ for Fig. \ref{t3}- \ref{t1}, $400$ for Fig. \ref{t0} - \ref{t-1}, and $800$ for Fig. \ref{t-2}.
The computation confirms that the time step for GKS is limited to be small when the cell Reynolds number is small, as shown in Fig. \ref{fig:cfl}.
The solution provided in UGKS is valid in all regimes from the kinetic $t \simeq \tau$ to the hydrodynamic one $t >> \tau$.
Both GKS and UGKS solutions converge in the hydrodynamic regime.
Since a much large cell size is used for the hydrodynamic solution for the case of $t >> \tau$ and $\Delta x >> l_{mfp}$,
the discontinuity cannot be well-resolved by both GKS and UGKS. A shock capturing approach is basically used for both schemes.
In order to give a more accurate physical representation, the cell size used in UGKS should depend on the flow physics to be resolved. In the highly dissipative region, a small cell
size is used for the capturing of non-equilibrium dynamics, and  in the smooth region a large cell size for the  hydrodynamic solution is accurate enough for its evolution.
The evolution solutions in different scales clearly indicate the usefulness of the direct modeling UGKS in comparison with the single scale NS solution.


\section{Conclusion}

The gas dynamics evolution  has an intrinsic multiple scale nature, which depends on the modeling scale relative to the particle mean free path.
In this paper, we present two physical descriptions for gas evolution.
One is the macroscopic based fluid element approach, i.e., the NS equations, and the other is the multiscale modeling algorithm UGKS.
This study presents the limitation of the macroscopic level modeling due to its fluid element assumption.
In the low Reynolds number limit, the NS approach
imposes severe time-step constraint, such as $\Delta t < {(\Delta x)}^2  / \nu$, for the capturing of flow evolution.
This time step limitation is purely an artificial one due to inappropriate macroscopic modeling for the microscopic scale physics.
In other words, the computational difficulties associated with NS solution at  low Reynolds number case
is from its physical inconsistency. For example, the fluid element assumption is still used in NS modeling to the cases where the particle penetration effect plays an important role
in the scale smaller than the hydrodynamic one.
For the direct modeling method, such as the UGKS, due to its capability to have a smooth transition  between the fluid element and particle free penetration mechanism
 with a variation of scales, the time step used in the computation is independent of Reynolds number, which is consistent
with the physical propagating speed in different regimes.
This study indicates that the direct modeling and computation for gas dynamics can provide an indispensable tool for the capturing of multiscale gas evolution.
The numerical computation is not necessarily to target on the exact solution of a specific governing equation, but to model the physical reality in the mesh size scale,
and construct the corresponding evolution model in such a scale.
The UGKS provides both equations and the evolution solution. It goes beyond the traditional numerical PDE principle for the computation.
The direct modeling methodology provides a new way for scientific computing, especially for the multiple scale transport, such as rarefied flow \cite{jiang}, radiative transfer \cite{sun}, phonon heat transfer \cite{guo},
and plasma physics \cite{liu-thesis}.
In the direct modeling scheme, we have no a fixed scale and a fixed  governing equation to be solved.
All well developed principles in numerical PDE, such as stability analysis, consistency, and the convergence, have to be reformulated under the direct modeling methodology.

\section*{Acknowledgement} %% The authors wish to thank the referees for the useful suggestions which improve this paper.
The current research  is supported by Hong Kong research grant council (16207715,16211014,620813) and  NSFC-91330203 and 91530319.

\end{document}